\documentclass[12pt, onecolumn, draftclsnofoot]{IEEEtran}%

\usepackage[T1]{fontenc}
\usepackage[utf8]{inputenc}
\usepackage{amsmath,amssymb,amsfonts,mathrsfs,bm}
\usepackage{mathtools}
\usepackage{amsthm}
\usepackage{nicefrac}
\usepackage[shortlabels]{enumitem}
\usepackage{graphicx}
\usepackage{epstopdf}
\usepackage{url}
\usepackage{colortbl}
\usepackage{booktabs}
\usepackage{multirow}
\usepackage[table,dvipsnames]{xcolor}
\usepackage[normalem]{ulem}
\usepackage{xparse}

\makeatletter
\@ifpackageloaded{natbib}{
	\relax
}{
	\usepackage{cite}
}
\makeatother

\usepackage{array}
\newcolumntype{L}[1]{>{\raggedright\let\newline\\\arraybackslash\hspace{0pt}}m{#1}}
\newcolumntype{C}[1]{>{\centering\let\newline\\\arraybackslash\hspace{0pt}}m{#1}}
\newcolumntype{R}[1]{>{\raggedleft\let\newline\\\arraybackslash\hspace{0pt}}m{#1}}

\makeatletter
\let\MYcaption\@makecaption
\makeatother
\usepackage[font=footnotesize]{subcaption}
\makeatletter
\let\@makecaption\MYcaption
\makeatother



\usepackage{glossaries}

\makeatletter
\let\oldgls\gls
\let\oldglspl\glspl

\newcommand\fussy@ifnextchar[3]{%
	\let\reserved@d=#1%
	\def\reserved@a{#2}%
	\def\reserved@b{#3}%
	\futurelet\@let@token\fussy@ifnch}
\def\fussy@ifnch{%
	\ifx\@let@token\reserved@d
		\let\reserved@c\reserved@a
	\else
		\let\reserved@c\reserved@b
	\fi
	\reserved@c}

\renewcommand{\gls}[1]{%
\oldgls{#1}\fussy@ifnextchar.{\@checkperiod}{\@}}
\renewcommand{\glspl}[1]{%
\oldglspl{#1}\fussy@ifnextchar.{\@checkperiod}{\@}}

\newcommand{\@checkperiod}[1]{%
	\ifnum\sfcode`\.=\spacefactor\else#1\fi
}
\makeatother

\newacronym{wrt}{w.r.t.}{with respect to}
\newacronym{RHS}{R.H.S.}{right-hand side}
\newacronym{LHS}{L.H.S.}{left-hand side}
\newacronym{iid}{i.i.d.}{independent and identically distributed}

\usepackage{float}

\ifx\notloadhyperref\undefined
	\ifx\loadbibentry\undefined
		\usepackage[hidelinks,hypertexnames=false]{hyperref}
	\else
		\usepackage{bibentry}
		\makeatletter\let\saved@bibitem\@bibitem\makeatother
		\usepackage[hidelinks,hypertexnames=false]{hyperref}
		\makeatletter\let\@bibitem\saved@bibitem\makeatother
	\fi
\else
	\ifx\loadbibentry\undefined
		\relax
	\else
		\usepackage{bibentry}
	\fi
\fi

\usepackage[capitalize]{cleveref}
\crefname{equation}{}{}
\Crefname{equation}{}{}
\crefname{claim}{claim}{claims}
\crefname{step}{step}{steps}
\crefname{line}{line}{lines}
\crefname{condition}{condition}{conditions}
\crefname{dmath}{}{}
\crefname{dseries}{}{}
\crefname{dgroup}{}{}

\crefname{Problem}{Problem}{Problems}
\crefformat{Problem}{Problem~(#2#1#3)}
\crefrangeformat{Problem}{Problems~(#3#1#4) to~(#5#2#6)}

\crefname{Theorem}{Theorem}{Theorems}
\crefname{Corollary}{Corollary}{Corollaries}
\crefname{Proposition}{Proposition}{Propositions}
\crefname{Lemma}{Lemma}{Lemmas}
\crefname{Definition}{Definition}{Definitions}
\crefname{Example}{Example}{Examples}
\crefname{Assumption}{Assumption}{Assumptions}
\crefname{Remark}{Remark}{Remarks}
\crefname{Rem}{Remark}{Remarks}
\crefname{remarks}{Remarks}{Remarks}
\crefname{Appendix}{Appendix}{Appendices}
\crefname{Supplement}{Supplement}{Supplements}
\crefname{Exercise}{Exercise}{Exercises}
\crefname{Theorem_A}{Theorem}{Theorems}
\crefname{Corollary_A}{Corollary}{Corollaries}
\crefname{Proposition_A}{Proposition}{Propositions}
\crefname{Lemma_A}{Lemma}{Lemmas}
\crefname{Definition_A}{Definition}{Definitions}

\usepackage{crossreftools}
\ifx\notloadhyperref\undefined
\else
	\relax
\fi

\usepackage{algorithm,algorithmic}

\ifx\loadbreqn\undefined
	\relax
\else
	\usepackage{breqn}
\fi

\ifx\renewtheorem\undefined
	\ifx\useTheoremCounter\undefined
		\newtheorem{Theorem}{Theorem}
		\newtheorem{Corollary}{Corollary}
		\newtheorem{Proposition}{Proposition}
		\newtheorem{Lemma}{Lemma}
	\else
		\newtheorem{Theorem}{Theorem}
		
		\newtheorem{Proposition}[theorem]{Proposition}
	\fi

	\newtheorem{Definition}{Definition}
	\newtheorem{Example}{Example}
	\newtheorem{Remark}{Remark}

\fi

\newcommand{\nn}{\nonumber\\ }

\theoremstyle{remark}

\theoremstyle{plain}

\newcommand{\qednew}{\nobreak \ifvmode \relax \else
		\ifdim\lastskip<1.5em \hskip-\lastskip
			\hskip1.5em plus0em minus0.5em \fi \nobreak
		\vrule height0.75em width0.5em depth0.25em\fi}

\NewDocumentCommand{\movedownsub}{e{^_}}{%
	\IfNoValueTF{#1}{%
		\IfNoValueF{#2}{^{}}
	}{%
		^{#1}
	}%
	\IfNoValueF{#2}{_{#2}}
}

\let\latexchi\chi
\RenewDocumentCommand{\chi}{}{\latexchi\movedownsub}

\newcommand{\Real}{\mathbb{R}}

\newcommand{\iu}{\mathfrak{i}\mkern1mu}

\newcommand{\calB}{\mathcal{B}}
\newcommand{\calC}{\mathcal{C}}

\newcommand{\calF}{\mathcal{F}}

\newcommand{\calU}{\mathcal{U}}

\newcommand{\bg}{\mathbf{g}}

\newcommand{\bI}{\mathbf{I}}

\newcommand{\bu}{\mathbf{u}}
\newcommand{\bU}{\mathbf{U}}

\newcommand{\bx}{\mathbf{x}}

\newcommand{\mbf}{\bm{f}}

\newcommand{\scB}{\mathscr{B}}

\newcommand{\scF}{\mathscr{F}}

\DeclareSymbolFont{bsfletters}{OT1}{cmss}{bx}{n}
\DeclareSymbolFont{ssfletters}{OT1}{cmss}{m}{n}
\DeclareMathSymbol{\bsfGamma}{0}{bsfletters}{'000}
\DeclareMathSymbol{\ssfGamma}{0}{ssfletters}{'000}
\DeclareMathSymbol{\bsfDelta}{0}{bsfletters}{'001}
\DeclareMathSymbol{\ssfDelta}{0}{ssfletters}{'001}
\DeclareMathSymbol{\bsfTheta}{0}{bsfletters}{'002}
\DeclareMathSymbol{\ssfTheta}{0}{ssfletters}{'002}
\DeclareMathSymbol{\bsfLambda}{0}{bsfletters}{'003}
\DeclareMathSymbol{\ssfLambda}{0}{ssfletters}{'003}
\DeclareMathSymbol{\bsfXi}{0}{bsfletters}{'004}
\DeclareMathSymbol{\ssfXi}{0}{ssfletters}{'004}
\DeclareMathSymbol{\bsfPi}{0}{bsfletters}{'005}
\DeclareMathSymbol{\ssfPi}{0}{ssfletters}{'005}
\DeclareMathSymbol{\bsfSigma}{0}{bsfletters}{'006}
\DeclareMathSymbol{\ssfSigma}{0}{ssfletters}{'006}
\DeclareMathSymbol{\bsfUpsilon}{0}{bsfletters}{'007}
\DeclareMathSymbol{\ssfUpsilon}{0}{ssfletters}{'007}
\DeclareMathSymbol{\bsfPhi}{0}{bsfletters}{'010}
\DeclareMathSymbol{\ssfPhi}{0}{ssfletters}{'010}
\DeclareMathSymbol{\bsfPsi}{0}{bsfletters}{'011}
\DeclareMathSymbol{\ssfPsi}{0}{ssfletters}{'011}
\DeclareMathSymbol{\bsfOmega}{0}{bsfletters}{'012}
\DeclareMathSymbol{\ssfOmega}{0}{ssfletters}{'012}

\DeclareMathOperator{\sinc}{sinc}

\DeclareMathOperator{\ima}{im}

\DeclarePairedDelimiter\parens{(}{)}

\DeclarePairedDelimiterX\ip[2]{\langle}{\rangle}{#1,#2}
\DeclarePairedDelimiterX\norm[1]{\lVert}{\rVert}{#1}
\DeclarePairedDelimiterXPP\col[1]{\operatorname{col}}{\{}{\}}{}{#1} 
\DeclarePairedDelimiterXPP\row[1]{\operatorname{row}}{\{}{\}}{}{#1} 
\DeclarePairedDelimiterXPP\erf[1]{\operatorname{erf}}{(}{)}{}{#1}
\DeclarePairedDelimiterXPP\erfc[1]{\operatorname{erfc}}{(}{)}{}{#1}
\DeclarePairedDelimiterXPP\op[2]{\operatorname{#1}}{(}{)}{}{#2} 

\newcommand{\setcomp}{^{\mathsf{c}}} 
\newcommand{\ud}{\,\mathrm{d}} 

\newcommand{\bzero}{\bm{0}}

\providecommand\given{}
\newcommand\SetSymbol[2][]{%
	\nonscript\, #1#2
	\allowbreak
	\nonscript\,
	\mathopen{}}

\DeclarePairedDelimiterX\Set[2]\{\}{%
\renewcommand\given{\SetSymbol[\delimsize]{#1}}
#2
}
\DeclarePairedDelimiterX\Setc[1]\{\}{%
\renewcommand\given{\SetSymbol{:}}
#1
}

\NewDocumentCommand\set{s o m}{%
	\IfBooleanTF#1%
	{\IfValueTF{#2}{\Set*{#2}{#3}}{\Setc*{#3}}}%
	{\IfValueTF{#2}{\Set{#2}{#3}}{\Setc{#3}}}%
}

\NewDocumentCommand{\evalat}{s O{\big} m m}{%
\IfBooleanTF{#1}
{{\left. #3 \right|_{#4}}}
{{#3#2|_{#4}}}%
}

\NewDocumentCommand \ifcond {m m} {%
	{#1} %
	\IfValueT{#2}{\, \middle|\, {#2}}%
}

\DeclareDocumentCommand \P {e{_} g >{\SplitArgument{ 1 }{ @| }}d() g } {%
	\mathbb{P}%
	\IfValueTF{#1}{_{#1}}
	{\IfValueT{#2}{_{#2}}}%
	\IfValueT{#3}{\left(\ifcond#3}%
	\IfValueT{#4}{\, \middle|\, {#4}}%
	\IfValueT{#3}{\right)}%
}

\DeclareDocumentCommand \E {e{_} g >{\SplitArgument{ 1 }{ @| }}o g } {%
\mathbb{E}%
\IfValueTF{#1}{_{#1}}
{\IfValueT{#2}{_{#2}}}%
\IfValueT{#3}{\left[\ifcond#3}%
\IfValueT{#4}{\, \middle|\, {#4}}%
\IfValueT{#3}{\right]}%
}

\let\oldforall\forall
\renewcommand{\forall}{\oldforall \, }

\let\oldexist\exists
\renewcommand{\exists}{\oldexist \, }

\graphicspath{{./Figures/}{./figures/}}
\newcommand{\includeCroppedPdf}[2][]{%
	\IfFileExists{./Figures/#2-crop.pdf}{}{%
		\immediate\write18{pdfcrop ./Figures/#2 ./Figures/#2-crop.pdf}}%
	\includegraphics[#1]{./Figures/#2-crop.pdf}}

\definecolor{gray90}{gray}{0.9}

\ifx\nohighlights\undefined

	\newcommand{\msout}[1]{\text{\color{green} \sout{\ensuremath{#1}}}}
	\newcommand{\del}[1]{{\color{green}\ifmmode \msout{#1}\else\sout{#1}\fi}}
\else

	\newcommand{\msout}[1]{#1}
	\newcommand{\del}[1]{#1}
\fi

\newcommand{\hhide}[1]{}

\ifx\diagnoselabel\undefined
	\relax
\else
	\makeatletter
	\def\@testdef #1#2#3{%
		\def\reserved@a{#3}\expandafter \ifx \csname #1@#2\endcsname
			\reserved@a  \else
			\typeout{^^Jlabel #2 changed:^^J%
				\meaning\reserved@a^^J%
				\expandafter\meaning\csname #1@#2\endcsname^^J}%
			\@tempswatrue \fi}
	\makeatother
\fi

\usepackage{tikz}
\usepackage{pgfplots}
\pgfplotsset{compat=1.5}
\newcommand{\BWset}{W_{\calB,\calC}}

\begin{document}
\title{Sampling Theory of Bandlimited Continuous-Time Graph Signals}

\author{Feng~Ji, Hui~Feng, Hang~Sheng and Wee~Peng~Tay,~\IEEEmembership{Senior Member,~IEEE}%
\thanks{F. Ji and W. P. Tay are supported by the Singapore Ministry of Education Academic Research Fund Tier 2 grant MOE2018-T2-2-019 and A*STAR under its RIE2020 Advanced Manufacturing and Engineering (AME) Industry Alignment Fund – Pre Positioning (IAF-PP) (Grant No. A19D6a0053).}%
\thanks{F. Ji and W. P. Tay are with the School of Electrical and Electronic Engineering, Nanyang Technological University, 639798, Singapore. H~Feng and H~Sheng are with School of Information Science and Technology, Fudan University, Shanghai, China.}
}

\maketitle

\begin{abstract}
  A continuous-time graph signal can be viewed as a time series of graph signals. It generalizes both the classical continuous-time signal and ordinary graph signal. Therefore, such a signal can be considered as a function on two domains: the graph domain and the time domain. In this paper, we consider the sampling theory of bandlimited continuous-time graph signals. To formulate the sampling problem, we need to consider the interaction between the graph and time domains. We describe an explicit procedure to determine a discrete sampling set for perfect signal recovery. Moreover, in analogous to the Nyquist-Shannon sampling theorem, we give an explicit formula for the minimal sample rate.
\end{abstract}

\section{Introduction}
Since its emergence, the theory and applications of graph signal processing (GSP) have rapidly developed \cite{Shu13, San13, San14, Gad14, Don16, Egi17, Sha17, Gra18, Ort18, Girault2018}. There are many works that generalize the basic GSP framework by extending the domain of application. One such direction is the time-vertex graph signal processing \cite{Gra18, Ji19}, with \cite{Ji19} contains one of the most general frameworks. Essentially, a signal $f(v,t)$ on a graph $G=(V,E)$ has both a graph component $f_t(\cdot) = f(\cdot, t)$ and a time component $f_v(\cdot) = f(v, \cdot)$. If the graph is a single vertex, then $f$ is nothing but a classical continuous-time signal on $\mathbb{R}$. On the other hand, for any snapshot $t$, $f_t$ is an ordinary graph signal.

Sampling theory is an important topic in signal processing. It permits signal recovery from signal values at a prescribed discrete set of points. In classical GSP theory, sampling has been studied extensively \cite{Aga13, Che15, Tsi15, Mar16, Anis2016}. The basic form of sampling problems amounts to choosing a suitable set of coordinates in a finite-dimensional vector space that is also a basis of a sparse subspace of graph signals.

Some works have been extended to the time-vertex signal processing framework. For example, in \cite{Gra18, Yu19}, sampling is considered for signals with finite discrete-time component, when signals belong to a finite-dimensional space. \cite{Ji19} considers signals with infinite-dimensional time components. It shows that for signal recovery, one can use asynchronous sampling, meaning samples can be chosen according to certain random procedures.

In classical signal processing, one of the fundamental results is the Nyquist-Shannon sampling theorem \cite{Sha49}. It gives both necessary and sufficient conditions for the size of a sampling set that permits perfect signal recovery. However, none of the above-mentioned works extend this important result to sampling for continuous-time graph signals. In this paper, we are going to offer a complete solution to this problem. By comparing GSP with the Nyquist-Shannon theory, we notice that finite dimensionality versus infinite dimensionality leads to principally distinct approaches. To study sampling for continuous-time graph signals, we need to reconcile the disparities between finiteness and infinitude. As the problem is more complicated, the answer is not as concise as the Nyquist-Shannon sampling theorem. It involves an inductive procedure and a series of statements with one generalizes the Nyquist-Shannon sampling theorem.

The rest of the paper is organized as follows. In \cref{sec:pro}, we define the notion of ``bandwidth'' for continuous-time graph signal, and formulate the sampling problem. At the end of \cref{sec:pro}, we summarize the main results of the paper. In \cref{sec:red}, we introduce the notion of ``simple GFT bandwidth''. We show the general sampling problem can be reduced to sampling for a signal space with simple GFT bandwidth in finitely many iterations. \cref{sec:simple} echoes \cref{sec:red} and discusses explicitly sampling for signal space with simple GFT bandwidth. The reduction step allows us to break up a signal space into smaller, more manageable pieces in terms sampling. In \cref{sec:adm}, we discuss how to put them back together for an overall sampling scheme. We demonstrate the procedures with an example in \cref{sec:eg} and conclude in \cref{sec:con}.

Notations. Let $\Real$ be the set of real numbers, $\Real_{+}$ be the set of non-negative real numbers and $\overline{\Real}_{+}=\Real_{+}\cup\{\infty\}$. We denote column and row vectors as well as matrices by boldface characters. Let $R$ and $C$ be row and column index subsets of a matrix $\bU$. The submatrix of $\bU$ corresponding to these rows and columns is denoted as $\bU_{R,C}$. Given a function $f\in L^2(V\times\Real)$ where $V$ is a finite set and $V'\subset V$, $\mbf_{V'}(t) \in \Real^{|V'|}$ is a column vector whose components are $f$ evaluated on $V'\times\set{t}$. The $v$-th component of a vector $\mbf$ is also denoted as $\mbf[v]$. The complement of a set $A \subset V$ is $A\setcomp = V\backslash A$. Let $\bI$ be the identity matrix and $[\cdot\, ;\, \cdot]$ denote vertical concatenation of matrices or vectors.

\section{Problem formulation} \label{sec:pro}

In this section, we formulate the sampling problem of bandlimited continuous-time graph signals. In addition, we give a glimpse of the main results we shall present in the paper.

Suppose $G=(V,E)$ is an undirected graph with vertex set $V$ and edge set $E$. Suppose $L$ is a symmetric graph shift operator, e.g., the adjacency matrix or the graph Laplacian of $G$. Let $\Lambda$ be the set of eigenvalues of $L$ also called \emph{graph frequencies}. Write $\bU$ for the matrix whose rows  $\set{\bu_{\lambda} \given \lambda\in \Lambda}$ are (transposed) orthonormal eigenvectors of $L$ associated with $\lambda\in \Lambda$. Given subsets of frequencies $\Lambda'\subset\Lambda$ and vertices $V'\subset V$, we use $\bU_{\Lambda', V'}$ to denote the submatrix of $\bU$ whose rows correspond to $\Lambda'$ and columns correspond to $V'$.

A \emph{continuous-time graph signal} $f(v,t)$ is a function in $L^2(V\times \mathbb{R})$, the space of square integrable functions on the domain $V\times \mathbb{R}$. Its restriction to each vertex $v\in V$ is denoted by $f_v(\cdot) = f(v,\cdot) \in L^2(\mathbb{R})$, intuitively understood as a signal in the ``time'' direction.

\begin{figure}[!htb]
  \centering
  \includegraphics[width=0.6\columnwidth]{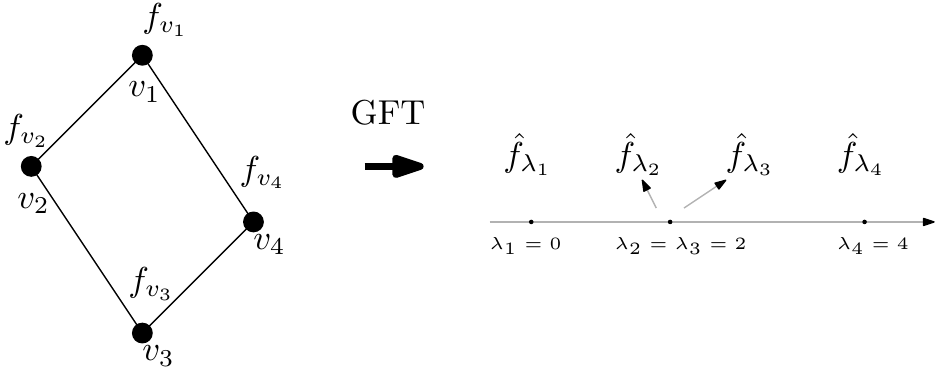}
  \caption{On the graph with $4$ vertices, each $f_{v_i}, 1\leq i\leq 4$ is a function associated with a vertex. On the other hand, their graph Fourier transforms $\{\widehat{f}_{\lambda_i}, 1\leq i\leq 4\}$ are associated with graph frequencies $\lambda_1,\lambda_2,\lambda_3,\lambda_4$. Nevertheless, all these functions belong to $L^2(\mathbb{R})$, and we can talk about their respective bandwidths.} \label{fig:cts5}.
\end{figure}

For each $t\in\Real$, $\mbf_V(t) = f(\cdot,t)\in\Real^{|V|}$ is a graph signal on $V$. For each frequency $\lambda\in \Lambda$, let
\begin{align}\label{eq:GFT}
  \widehat{f}_{\lambda}(t) = \bu_{\lambda} \mbf_V(t)
\end{align}
be the graph Fourier transform (GFT) \cite{Shu13} of $\mbf_V(t)$ at frequency $\lambda$. Both $f_v$ and $\widehat{f}_{\lambda}$ belong to $L^2(\mathbb{R})$ (see \cref{fig:cts5} for an illustration), and we can define their respective bandwidths. Let $\scF : L^2(\Real) \mapsto L^2(\Real)$ be the Fourier transform operator:\footnote{In this paper, we consider Fourier transform only on $L^2(\Real)$.} for any $g\in L^2(\Real)$,
  \begin{align}
    \scF{g}(\omega) = \int_{\Real} g(x) e^{-\iu \omega x} \ud x,
  \end{align}
  where $\iu = \sqrt{-1}$. The \emph{bandwidth} of $g$ is given by $\sup\set{|\omega| \given |\scF{g}(\omega)|>0} \in \overline{\Real}_{+}$. The signal $g$ is said to be \emph{bandlimited} if its bandwidth is finite. We say that a set of signals has bandwidth $b$ if every signal in the set has bandwidth bounded by $b$.
  
  From the generalized GSP theory of \cite{Ji19}, a continuous-time graph signal $f\in L^2(V\times \Real)$ is characterized by its joint $\calF$-transform, which is given by $\set{\scF{\widehat{f}_{\lambda}} \given \lambda\in\Lambda}$. As physical sampling of the signal $f$ takes place over the domain $V\times\Real$, i.e., on the signals $\set{f_v \given v\in V}$, we are interested in the interplay between the bandwidths of $\set{f_v \given v\in V}$ and $\set{\widehat{f}_{\lambda} \given \lambda\in\Lambda}$. We therefore have the following definition.

\begin{Definition}
  Let $\calB = \set*{\calB[v] \in \overline{\Real}_{+} \given v \in V}$ and $\calC = \set*{\calC[\lambda] \in \overline{\Real}_{+} \given \lambda \in \Lambda}$ be collections of bandwidths. A continuous-time graph signal $f\in L^2(V\times\Real)$ has bandwidths bounded by $(\calB,\calC)$ if the bandwidth of $f_v$ is bounded by $\calB[v]$ for each $v\in V$, and the bandwidth of $\widehat{f}_{\lambda}$ is bounded by $\calC[\lambda]$ for each $\lambda \in \Lambda$.

  The vector space $\BWset$ consists of continuous-time graph signals $f$ whose bandwidths are bounded by $(\calB,\calC)$. We say that $\BWset$ is \emph{uniformly bandlimited} if for any $f \in \BWset$ and $v\in V$, $f_v$ has finite bandwidth, uniformly bounded independent of $f$.
\end{Definition}

\begin{Remark}\label{rem:bw}
  We first remark that an infinite bandwidth means that no restriction is imposed for the signal at a particular vertex or graph frequency. On the other hand, for a signal in $L^2(\Real)$, a bandwidth of zero means that the signal is a constant value almost everywhere (a.e.). Since we have assumed signals to be in $L^2(V\times\Real)$, this implies that the signal is the constant zero a.e. Furthermore, from \cref{eq:GFT}, if a GFT signal $\widehat{f}_{\lambda}=0$, it implies that the subset of vertex signals $\set{f_v\given \bu_{\lambda}[v] \ne 0}$ are linearly dependent.
\end{Remark}

If $\max_{v\in V} \calB[v] <\infty$, then the space $\BWset$ is by definition uniformly bandlimited. On the other hand, if $\calB[v]=\infty$ for some $v\in V$, it is still possible for $\BWset$ to be uniformly bandlimited, depending on the bandwidth constraints imposed by $\calC$. We now give conditions on $\calB$ and $\calC$ for this case.

\begin{Lemma} \label{lem:lvb}
  For a collection $\calB$, let $V_{\infty}$ be the subset of $v \in V$ such that $\calB[v]=\infty$. Suppose that $V_\infty \ne \emptyset$. Then $\BWset$ is uniformly bandlimited if and only if there is a subset $\Lambda'\subset\Lambda$ of size $|V_{\infty}|$ such that:
  \begin{enumerate}[(a)]
    \item the matrix $\bU_{\Lambda',V_{\infty}}$ is invertible, and
    \item for all $\lambda \in \Lambda'$, $\calC[\lambda] < \infty$.
  \end{enumerate}
\end{Lemma}
\begin{IEEEproof}
  Suppose $\BWset$ is uniformly bandlimited and every subset $\Lambda'\subset\Lambda$ with invertible $\bU_{\Lambda', V_{\infty}}$ contains some $\lambda$ such that $\calC[\lambda] = \infty$. Consider the matrix $\bU_{\Lambda, V_{\infty}}$, which has shape $|\Lambda|\times |V_\infty|$. By the assumption, there are at most $|V_\infty|-1$ independent rows in $\bU_{\Lambda, V_{\infty}}$ whose corresponding graph frequencies $\lambda$ satisfy $\calC[\lambda] < \infty$. Denote one such maximal collection of $\lambda$'s (with corresponding rows independent) by $\Lambda_0$. This means that any row of $\bU_{\Lambda, V_{\infty}}$ corresponding to a graph frequency $\lambda'$ with $\calC[\lambda']<\infty$ is a linear combination of the rows corresponding to $\Lambda_0$. Now $\bU_{\Lambda_0, V_{\infty}}$ has more columns than rows, and we can construct a continuous-time graph signal $f$ such that
  \begin{enumerate}[i)]
    \item $f_v = 0$ for $v\notin V_{\infty}$;
    \item $f_v$ for some $v\in V_{\infty}$ has arbitrary large bandlimit; and
          \item\label{it:GFT0} $\bU_{\Lambda_0, V_{\infty}} \mbf_{V_\infty}(t)=0$ for all $t\in\Real$.
  \end{enumerate}
 By construction, $f_v$, $v\in V$ have bandwidths bounded by $\calB$. We next show that $\widehat{f}_\lambda$, $\lambda\in\Lambda$ have bandwidths bounded by $\calC$. From condition~\ref{it:GFT0} and \cref{eq:GFT}, $\widehat{f}_\lambda = 0$ for $\lambda\in\Lambda_0$. As a consequence, $\widehat{f}_{\lambda'}=0$ for each $\lambda'$ with $\calC[\lambda']<\infty$, as $\widehat{f}_{\lambda'}$ is a linear combination of the functions $\set{\widehat{f}_{\lambda} \given \lambda\in\Lambda_0}$. Therefore, $f\in\BWset$ and this contradicts the assumption that $\BWset$ is uniformly bandlimited.

  We now prove the other direction. Since $\bU_{\Lambda', V_{\infty}}$ is invertible, we have for each $t\in\Real$,
    \begin{align}
       & \bU_{\Lambda',V_\infty}\mbf_{V_\infty}(t) + \bU_{\Lambda',V_\infty\setcomp}\mbf_{V_\infty\setcomp}(t) = \widehat{\mbf}_{\Lambda'}(t)                 \\
       & \mbf_{V_\infty}(t) = \bU_{\Lambda',V_\infty}^{-1}\parens*{\widehat{\mbf}_{\Lambda'}(t) - \bU_{\Lambda',V_\infty\setcomp}\mbf_{V_\infty\setcomp}(t)}.
    \end{align}%
  Therefore, for each $v \in V_{\infty}$, $f_v$ is a linear combination of functions in $\set{f_{v'}\given v'\in V_{\infty}\setcomp} \cup \set{\widehat{f}_{\lambda}\given \lambda\in \Lambda'}$. The bandwidth of $f_v$ is thus bounded by
  \begin{align}\label{bwupper}
    \max\set*{\max_{v'\in V_{\infty}\setcomp} \calB[v'], \max_{\lambda \in \Lambda'} \calC[\lambda]},
  \end{align}
  which is independent of $f$. The proof is now complete.
\end{IEEEproof}

As a simple preprocessing step, we may modify $(\calB, \calC)$ if necessary such that $\calB[v] < \infty$ for each $v\in V$, which we \emph{assume for the rest of the paper}. To do this, by \cref{lem:lvb}, we first identify $V_{\infty}$ and the associated $\Lambda'$ in the statement of \cref{lem:lvb}. Then we replace each $\calB[v]$, $v\in V_{\infty}$, which is $\infty$, by \cref{bwupper}. Suppose $\BWset$ is known to be uniformly bandlimited. If $V_{\infty}$ and $V$ are too large such that it is intractable to find $\Lambda'$, we can just replace $\calB[v]$, $v\in V_{\infty}$ by $\max\{\max_{v\notin V_{\infty}}\calB[v],\max_{\lambda\in \Lambda}\calC[\lambda]\}$.


\begin{Example} \label{eg:sgi}
  \begin{enumerate}[(a)]
    \item Suppose $G=(V=\{v\}, E=\emptyset)$ is the trivial graph with a single vertex. Then it is sufficient to specify $\calB=\{\calB[v]\}$. By the Nyquist-Shannon sampling theorem, sampling at  evenly spaced points on $\Real$ at the rate $r = 2\calB[v]$, allows one to uniquely recover any signal in $\BWset$. Moreover, the rate $r$ is minimal. This is a result we will generalize.

          \item\label{it:G2} For the simplest nontrivial graph, let $G = (V=\{v_1,v_2\}, E=\set{(v_1,v_2)})$ be the graph with $2$ vertices connected by an edge. Let $L$ be the graph Laplacian. The graph frequencies are $\lambda_1=0$ and $\lambda_2=2$ with corresponding eigenvectors $\bu_{\lambda_1}=(1/\sqrt{2},1/\sqrt{2})$ and $\bu_{\lambda_2}=(1/\sqrt{2},-1/\sqrt{2})$. Consider $\calB = \{\calB[v_1], \calB[v_2]\}$ and $\calC = \{\calC[\lambda_1]=0, \calC[\lambda_2]=\infty\}$. For a signal $f\in \BWset$, the bandwidth constraint $\calC[\lambda_1]=0$ implies that $\widehat{f}_{\lambda_1}(t) = \bu_{\lambda_1}\mbf_V(t)=0$ for a.e. $t\in\Real$ (cf.\ \cref{rem:bw}), which enforces $f_{v_1}= -f_{v_2}$ a.e. This means that the signal at either at $v_1$ or $v_2$ determines the entire signal $f$. Therefore, to recover any signal of $\BWset$, one may sample either at rate $2\calB[v_1]$ along the vertex $v_1$ or $2\calB[v_2]$ along the vertex $v_2$, and $2\min\{\calB[v_1],\calB[v_2]\}$ is an upper bound of the minimal sample rate. However, the situation is more subtle if $0<\calC[\lambda_1]<\infty$, as this constraint does not lead to any simple algebraic identity. A general characterization will be provided in this paper.
  \end{enumerate}
\end{Example}

Our goal is to develop a sampling theory for $\BWset$ for general $\calB$ and $\calC$, i.e., we want to determine the minimal sampling rate required to recover any signal in $\BWset$ as well as a feasible sampling and recovery procedure. A ``sampling problem'' refers to the construction of a sample set with the minimal sampling rate to achieve perfect recovery of a signal from the samples. Our main results and exposition are summarized as follows.
\begin{enumerate}[(R.1)]
  \item\label{it:tia} (\cref{sec:red}) There is a finite filtration of subspaces
  \begin{align}\label{filtration}
    W_0 \subset \ldots W_i \subset \ldots \subset W_k=\BWset
  \end{align}
  for some $k <\infty$ such that the following holds: (a) the quotient space (see below) $W_{i}/W_{i-1}$ for each $i=1,\ldots,k$ can be identified with a space of bandlimited signals in $L^2(\mathbb{R})$; and (b) $W_0 = W_{\calB, \calC_0}$ where each $c\in\calC_0$ is either $0$ or $\infty$ (cf.\ \cref{eg:sgi}).

  \item\label{it:sfu} (\cref{sec:red}) Since each $W_{i}/W_{i-1}$, $1\leq i \leq k$, can be identified with a space of bandlimited signals in $L^2(\mathbb{R})$, sampling for $W_{i}/W_{i-1}$ can be performed using the classical theory of Nyquist-Shannon.

  \item\label{it:srf} (\cref{sec:simple}) We can write $W_0 = W_{\calB, \calC_0}$ such that the minimal sampling rate for $W_0$ can be computed explicitly by $2\sum_{v\in V_0} \calB[v]$ for a suitably chosen set of vertices $V_0 \subset V$.

  \item\label{it:sad} (\cref{sec:adm}) Under favorable conditions, we have a formula for optimal sample rate for a general $\BWset$ with an explicit sampling strategy for the rate.  
\end{enumerate}

We end this section by giving intuitions on how these results are used in sampling. We adopt an inductive procedure. First recall that suppose $H_1$ is a Hilbert space and $H_2$ is a closed subspace. Then the \emph{quotient space} $H_1/H_2$ are equivalence classes $[h]$ of elements of $H_1$, with $h_1, h_2 \in H_1$ are equivalent if $h_1-h_2 \in H_2$. For a class $[h] \in H_1/H_2$, its norm is the infimum of the $H_1$-norm of $h_1$ where $h_1 \in H_1$ belongs to the same equivalence class of $h$. This makes $H_1/H_2$ a Hilbert space and we have the (linear) quotient map $H_1\to H_1/H_2, h\mapsto [h]$. Its kernel is the space $H_2$.

Back to our situation, by \ref{it:srf}, we can perform sampling for $W_0$. For $W_1$, we consider the sequence of maps \begin{align*}W_0 \xrightarrow{\phi_1} W_1 \xrightarrow{\phi_2} W_1/W_0,\end{align*} where $\phi_1$ is the inclusion map and $\phi_2$ is the natural map to the quotient space. By \ref{it:sfu}, we have a sampling method for $W_1/W_0$. As a result, we know how to perform sampling for the spaces at the two ends of the sequence of maps. The next step is to sample for $W_1$ by combining sampling results for $W_0$ and $W_1/W_0$, under favorable conditions (c.f.\ \cref{sec:adm}). In particular, we shall see that the sample rate for the space in the middle of the sequence, i.e., $W_1$, is the sum of the sample rates for the spaces at the two ends. The same procedure can then be repeated for $W_i$, $i\geq 2$ until we finally sample for $\BWset$. Details are provided in \cref{sec:adm}. 

\section{Reduction to simple GFT bandwidths} \label{sec:red}

In this section, we show how the space $\BWset$ can be decomposed into simpler subspaces and finally reduced to a space $W_0$ of signals whose vertex signals can be sampled using a procedure we develop in the next \cref{sec:red}.

The following definition is motivated by \cref{eg:sgi}.

\begin{Definition}\label{def:simple}
  The space $\BWset$ is said to have \emph{simple GFT bandwidths} if $\calC[\lambda] \in \{0,\infty\}$ for all $\calC[\lambda]\in\calC$.
\end{Definition}

For a graph frequency $\lambda\in\Lambda$, if $\calC[\lambda]\in\calC$ is zero, then from \cref{eq:GFT}, we see that the subset of vertex signals $\set{f_v\given \bu_{\lambda}[v] \ne 0}$ are linearly dependent in the vector space $\BWset$. On the other hand, if $\calC[\lambda]=\infty$, the vertex signals $\set{f_v\given \bu_{\lambda}[v] \ne 0}$ are independent of each other. Intuitively, vertex signals in the latter set with $\calC[\lambda]=\infty$ can be sampled independently while vertex signals in the former set with $\calC[\lambda]=0$ can be ``reconstructed'' from the latter set. Therefore, a signal belonging to a space $\BWset$ with simple GFT bandwidths can be recovered through sampling under the classical Nyquist-Shannon theory. We give more details in \cref{sec:simple}.

\subsection{The reduction step}

To obtain the desired filtration $W_0 \subset \ldots W_i\subset \ldots \subset W_k=\BWset$, the strategy is to gradually modify $\calC$ so that we end up with simple GFT bandwidths. In the following, we show how to perform the reduction step on $W_k=\BWset$ to obtain $W_{k-1}$. The procedure can then be performed inductively until $W_0$ with simple GFT bandwidths is reached.

Consider a signal space $W_k=\BWset$. Let $\lambda^* \in \Lambda$ be the graph frequency such that 
\begin{align}
  \calC[\lambda^*] = \min_{\lambda \in \Lambda, 0<\calC[\lambda]<\infty} \calC[\lambda].
\end{align}
To reduce $\calC$ to a ``simpler'' set, we find a subspace in $\BWset$ whose signals have GFT bandwidth at $\lambda^*$ being zero. This is naturally given by the kernel of the following linear map induced by the GFT at frequency $\lambda^*$:
\begin{align}\label{GFTmap}
  \begin{aligned}
    \alpha_{\lambda^*} : & \BWset \mapsto L^2(\mathbb{R}),                                                 \\
                         & f(\cdot) \mapsto \widehat{f}_{\lambda^*}(\cdot) = \bu_{\lambda^*}\mbf_V(\cdot).
  \end{aligned}
\end{align}
Then, the following result follows immediately.
\begin{Lemma}
  For any given space $\BWset$, the kernel $\ker(\alpha_{\lambda^*}) = W_{\calB, \calC'}$, where $\calC$ and $\calC'$ differ only at $\lambda^*$ with $\calC'[\lambda^*] = 0$.
\end{Lemma}

To obtain the finite filtration as claimed in \ref{it:tia}, we set $W_{k-1}=\ker(\alpha_{\lambda^*})$, and repeat the procedure on $W_{k-1}$. In each iteration, there is exactly one $\lambda^*$ whose associated bandwidth changes from a positive value to $0$. Therefore, the procedure terminates in finitely many steps.

Any signal $f\in W_k$ can be written as $f = f_1 + f_2$, where $f_1 \in W_{k-1}$ and $f_2$ belongs to the complement space (\gls{wrt} the direct sum) of $W_{k-1}$. From the first isomorphism theorem of, this complement is isomorphic to $W_k/W_{k-1}$, which is also isomorphic to the image $\ima(\alpha_{\lambda^*})$. In \cref{sec:simple,sec:adm}, we show how $f_1$ can be sampled and recovered. We now focus on the sampling of $f_2$, which is used in \cref{sec:adm} to recover $f$. We introduce the following notion of uniqueness sets.

\begin{Definition} \label{defn:fps}
  For a graph frequency subset $\Lambda' \subset \Lambda$, a vertex set $V' \subset V$ such that $|V'|+|\Lambda'|=|V|$ and $\bU_{\Lambda', {V'}\setcomp}$ is invertible is called a uniqueness set \gls{wrt} $\Lambda'$. The collection of uniqueness sets \gls{wrt} $\Lambda'$ is denoted as $\calU(\Lambda')$. If $\Lambda_0 = \emptyset$, then we set $\calU(\Lambda')=\{V\}$.
\end{Definition}

The term ``uniqueness set'' comes from \cite{Pesenson2008}. To see the reason for this terminology, let
\begin{align}\label{Lambda0}
  \Lambda_0 = \Lambda_0(\calC) := \set{\lambda\in\Lambda\given \calC[\lambda]=0}.
\end{align}
For simplicity, we use the notation $\Lambda_0$ when it is clear from the context what is the bandwidth collection $\calC$. Consider a $V_0 \in \calU(\Lambda_0)$. From \cref{rem:bw}, we have for all $t\in\Real$,
\begin{align} \label{eq:bml}
   & \bU_{\Lambda_0, V}\mbf_V(t)=0,      \nn        
   & \bU_{\Lambda_0, V_0}\mbf_{V_0}(t) + \bU_{\Lambda_0,V_0\setcomp}\mbf_{V_0\setcomp}(t) = \bzero, \nn
   & \mbf_{V_0\setcomp}(t) = -\bU_{\Lambda_0,V_0\setcomp}^{-1} \bU_{\Lambda_0, V_0}\mbf_{V_0}(t), \nn
   & \mbf_V(t) = [\bI\, ;\, -\bU_{\Lambda_0,V_0\setcomp}^{-1} \bU_{\Lambda_0, V_0}] \mbf_{V_0}(t).
\end{align}
Therefore, $\mbf_V(t)$ is determined by $\mbf_{V_0}(t)$, the signals on the vertex subset $V_0$. Taking inner product with $\bu_{\lambda^*}$, we have
\begin{align} \label{eq:fuf}
  \alpha_{\lambda^*}(f)(t) & = \bx_{V_0}\mbf_{V_0}(t),
\end{align}
where
\begin{align}\label{bxV}
  \bx_{V_0} & = \bu_{\lambda^*}[\bI\, ;\, -\bU_{\Lambda_0,V_0\setcomp}^{-1} \bU_{\Lambda_0, V_0}] \in\Real^{1\times|V_0|}.
\end{align}
Therefore, whether $\alpha_{\lambda^*}(f)$ takes contribution from $f_v$, $v\in V_0$, or not depends on whether $\bx_{V_0}[v]$ is $0$ or not. For $V_0 \in \calU(\Lambda_0)$, set 
\begin{align*}
  b_{V_0} = \max \set{\calB[v] \given \bx_{V_0}[v] \ne 0}
\end{align*}
and
\begin{align}
  b_{\lambda^*} = \min\set*{\min_{V_0 \in \calU(\Lambda_0)}b_{V_0}, \calC[\lambda^*]}.
\end{align}

\begin{Proposition} \label{prop:tio}
  Consider a space $W_k=\BWset$. Let $W_{k-1}=\ker(\alpha_{\lambda^*}) \subset W_k$. Then, the image $\ima(\alpha_{\lambda^*}) \cong W_k\backslash W_{k-1}$ is the subspace of $L^2(\mathbb{R})$ with bandwidth $b_{\lambda^*}$, which is tight (i.e., there exists $f\in\BWset$ such that the bandwidth of $\alpha_{\lambda^*}(f)$ is exactly $b_{\lambda^*}$).
\end{Proposition}
\begin{IEEEproof}
  Recall that $\Lambda_0=\set{\lambda\in\Lambda\given \calC[\lambda]=0}$ in \cref{Lambda0}. For any $V_0 \in \calU(\Lambda_0)$, $\alpha_{\lambda^*}(f)$ is a linear combination of $f_v, v\in V_0$. Therefore, its bandwidth is bounded by $b_{\lambda^*}$. We only need to show the bound $b_{\lambda^*}$ is tight. To do so, it suffices to display an $f$ such that the bandwidth of $\alpha_{\lambda^*}(f)$ is exactly $b_{\lambda^*}$.

  We first consider the case that $\calC[\lambda^*]> b_{\lambda^*}$. Let $V_0 \in \calU(\Lambda_0)$ minimize $b_{V_0}$. We fix $v^* \in V_0$ such that $\bx_{V_0}[v^*]\neq 0$ and $b_{v^*}= b_{\lambda^*}$. For $v^*$, we choose a $g_{v^*}$ whose bandwidth is exactly $b_{\lambda^*}$. For other $v\in V_0$, we let $g_v=0$. We want to show that:
  \begin{enumerate}[(1)]
    \item There is an $f \in W_{\calB, \calC}$ such that $f_v= g_v, v\in V_0$.

    \item The bandwidth of $\alpha_{\lambda^*}(f)$ is $b_{\lambda^*}$.
  \end{enumerate}
  As $V_0\in \calU(\Lambda_0)$, the matrix $\bU_1 = \bU_{\Lambda_0, V_0\setcomp}$ is invertible. Denote by $\bg$ the vector $(g_v)_{v\in V_0}$. For each $v \notin V_0$, $f_v$ is the $v$-component of  $-\bU_1^{-1}\bU_{\Lambda_0, V_0}\bg$ (c.f.\ third line of (\ref{eq:bml})), whose bandwidth is bounded by $b_{v^*}$. If the bandwidth of $f_v$ is strictly larger than $\calB[v]$, then $\calB[v] < b_{v^*}$. Moreover, as $f_v\neq 0$, the column vector $\bU_{\Lambda_0,v}$ is independent of the column vectors of $\bU_{\Lambda_0, V_0\setcomp\backslash \{v^*\}}$. Otherwise, $f_v$ is a linear combination of $g_{v'}=0, v'\in V_0\backslash \{v\}$, which is $0$. Hence, we may replace $v^*$ by $v$ in forming $V_0' \in \calU(\Lambda_0)$. But $\calB[v] < b_{ v^*}$ and this means we can reduce further $b_{V_0}$ (after processing all such $v^*$), which is a contradiction to our choice of $V_0$.

  By our choice of $\lambda^*$, for $\lambda \notin \Lambda_0$, $\calC[\lambda]\geq b_{\lambda^*}$. As we have seen in \cref{eq:fuf}, $\widehat{f}_{\lambda}$ is linear in $g_v$, whose bandwidth is bounded by $b_{\lambda^*}$, which is in turn bounded by $\calC[\lambda]$. Therefore, $f \in \BWset$ as claimed in part (1). As $\bx_{V_0}[v^*]\neq 0$ and $\alpha_{\lambda^*}(f) = \bx_{V_0}[v^*]g_{v^*}$, claim (2) holds.

  The case for $\calC[\lambda]= b_{\lambda^*}$ is similar. We only need to modify $g_{v^*}$ so that its bandlimit is $b_{\lambda^*}$.
\end{IEEEproof}

\subsection{Sampling strategy} \label{sec:str}

As we mentioned earlier, we can repeatedly apply the proposition to obtain the desired filtration $W_0 \subset \ldots \subset W_k$ in finitely many steps. We can also extract more information from the proof of \cref{prop:tio} on explicit sampling strategy, which we describe now. 

\begin{figure}[!htb]
  \centering
  \includegraphics[width=0.8\columnwidth]{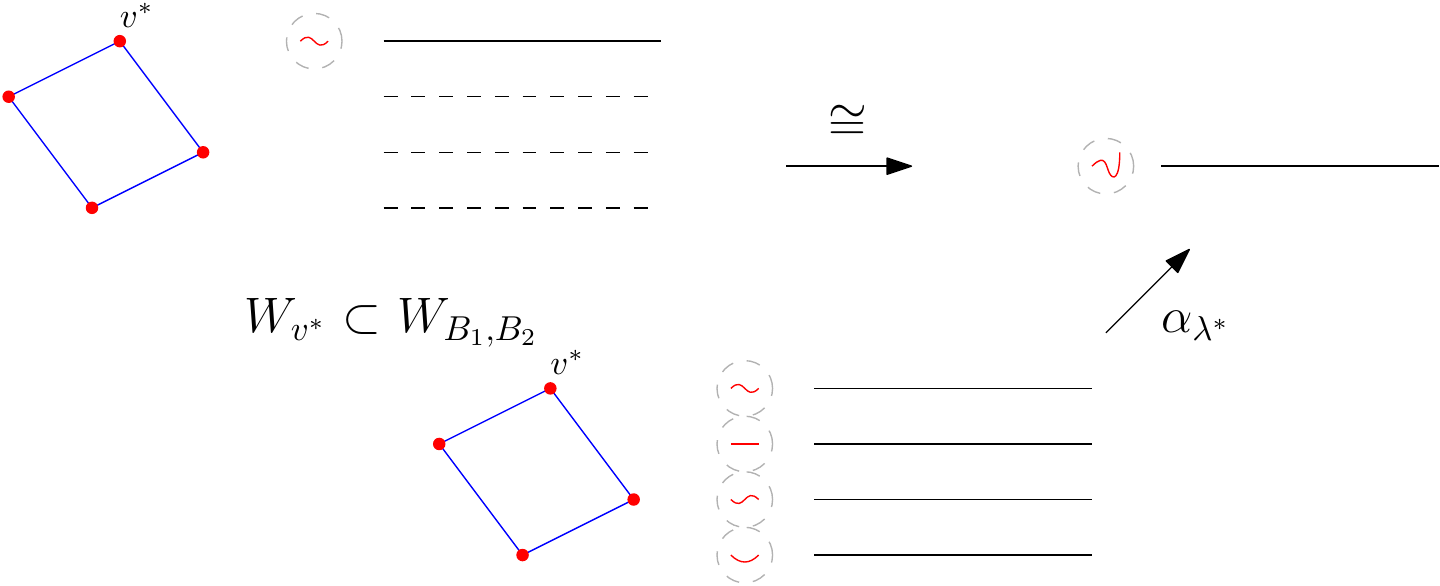}
  \caption{This figure illustrates that there is a subspace $W_{v^*}$ that is isomorphic to $\alpha_{\lambda^*}(\BWset) \subset L^2(\mathbb{R})$. For each function $f\in \alpha_{\lambda^*}(\BWset)$, to find a signal in its pre-image, we only need to sample along $v^*$ and recover a function $f_{v^*}$. The functions along the other vertices (dotted lines) are uniquely determined by $f_{v^*}$.} \label{fig:cts4}
\end{figure}

Consider a $V_0\in \calU(\Lambda_0)$ and $v^*\in V_0$. Let $W_{v^*}$ be the subspace of $\BWset$ consisting of $f$ such that $f_v=0$ for all $v\in V_0\backslash \{v^*\}$. We observe that there is a $V_0$ and $v^*\in V_0$ such that $\alpha_{\lambda^*}$ restricts to an isomorphism $W_{v^*} \to \alpha_{\lambda^*}(\BWset) \subset L^2(\mathbb{R})$. By \cref{prop:tio}, the bandwidth of $\alpha_{\lambda^*}(\BWset)$ is $b_{\lambda^*}$. Hence, for recovery of signals in $W_{v^*}$, it suffices to sample along $v^*$ with rate $2b_{\lambda^*}$. An illustration is given in \cref{fig:cts4}.

The above sampling strategy requires $v^*$ to satisfy conditions such as $\calB[v^*]= b_{\lambda^*}$. This can be restrictive in applications. Let us extract the essential ingredients. We notice that we need to find $v\in V_0\subset \calU(\Lambda_0)$ such that if we sample along $v$ and set $0$ along any other nodes in $V_0$, we can recover a signal $f$. The choice of $V_0$ is to ensure that $f$ does not violate the bandwidth condition on any of the nodes in $V_0\setcomp$. Put this formally, we have 

For any given $f$ in the pre-image of $\alpha_{\lambda^*}$, one may sample with rate $2b_{\lambda^*}$ at any vertex $v$ such that 
\begin{enumerate}[(a)]
  \item $v \in V_0$ for some $V_0 \subset \calU(\Lambda_0)$ and $\bx_{V_0}[v]\ne0$;
  \item $\calB[v] \geq b_{\lambda^*}$; and 
  \item $\calB[v'] \geq b_{\lambda^*}$ for each $v' \notin V_0$ and $f_{v'}$ cannot be expressed as a linear combination of $\set{f_{v''}\given v''\in V_0\backslash \{v\}}$ (cf.\ \cref{lem:fvv} in \cref{sec:simple}).
\end{enumerate}

Suppose we make observation at the samples along $v$, we can perform the following recovery steps:
\begin{enumerate}[(a)] 
\item \label{it:rau} Recover a unique $f_{v}$ with bandwidth $\calB[v]$ by the classical Nyquist-Shannon sampling theorem. 
\item For $v' \in V_0\backslash \{v\}$, we choose $f_{v'}=0$. 
\item \label{it:avc} As $V_0\in \calU(\Lambda_0)$, these functions together determines a unique $f\in \BWset$ by taking linear combinations according to \cref{eq:bml}.
\end{enumerate}

All $f$ constructed in this way forms a subspace of $\BWset$ isomorphic to the image of $\alpha_{\lambda^*}$. The discussions shall be used in \cref{sec:adm} below.

\section{Sampling for simple GFT bandwidth} \label{sec:simple}

In this section, we study the problem of sampling a signal in $\BWset$ with simple GFT bandwidth. We first define sample rate for a countable discrete subset of $V\times \mathbb{R}$.

\begin{Definition}
  Suppose $S$ is a countable discrete subset of $V\times \mathbb{R}$. As a sampling set, its \emph{sample rate} $r(S)$ is defined as $$r(S) = \limsup_{t\to \infty}\frac{|S\cap (V\times [-t,t])|}{2t}.$$
\end{Definition}

In this paper, we shall work with the case that $S\cap (V\times [k/r(S), (k+1)/r(S)]), k\in \mathbb{Z}$ is a constant for $|k|$ sufficiently large.

\begin{Definition}
  We say that $\BWset$ is \emph{tight} or $\calB$ is \emph{tight} \gls{wrt}\ $\calC$ if for each $v\in V$, there is an $f \in \BWset$ such that the bandwidth of $f_v$ is $\calB[v]$.
\end{Definition}

Intuitively, the tightness condition requires that none of the numbers $\calB[v]$ in $\calB$ can be reduced further.

\begin{Example}
  Consider the graph $G$ in \cref{eg:sgi}\ref{it:G2} with two vertices $v_1$ and $v_2$ connected by an edge of unit length. If $\calC[\lambda]=0$ for $\lambda=0$, i.e., $\bu_\lambda=(1/\sqrt{2},1/\sqrt{2})$, then $\calB=\{\calB[v_1], \calB[v_2]\}$ is tight \gls{wrt}\ $\calC$ if and only if $\calB[v_1]=\calB[v_2]$.
\end{Example}

\begin{Definition} \label{defn:lvba}
Let $V'\subset V$ be a subset of vertices and consider $\Lambda_0$ in \cref{Lambda0}. We say that a vertex $v \in {V'}\setcomp$ is $\Lambda_0$-dependent on $V'$ if: all vectors, in the orthogonal complement (in $\mathbb{R}^{|{V'}\setcomp|}$) of the rows of $\bU_{\Lambda_0, {V'}\setcomp}$, have zero $v$-component. 
\end{Definition}

The above notion of vertex dependency is motivated by the following lemma. In particular, if $V_0\in\calU(\Lambda_0)$ is a uniqueness set \gls{wrt} $\Lambda_0$, then any $v\in V_0\setcomp$ is $\Lambda_0$-dependent on $V_0$. To verify the condition, it suffices to check the condition on any basis of the orthogonal complement. 

\begin{figure}[!htb]
    \centering
    \includegraphics[width=0.4\columnwidth]{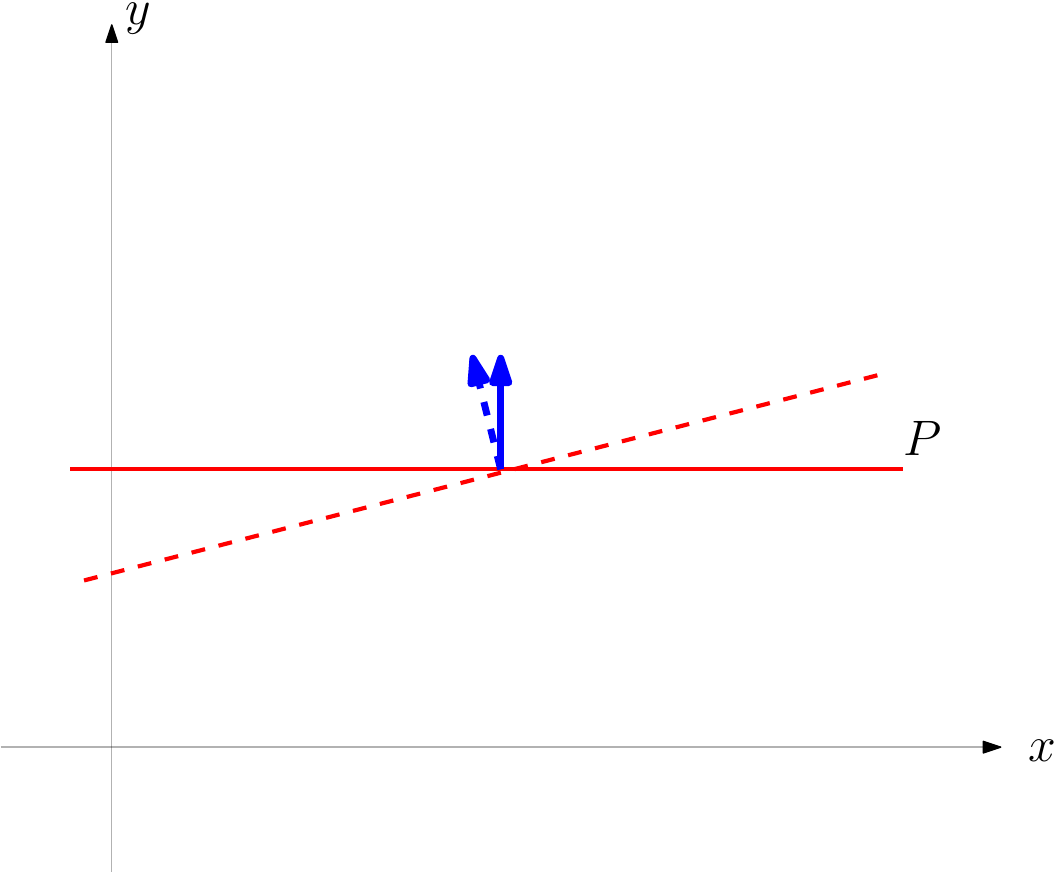}
    \caption{A line $P$, as a hyperplane in $\mathbb{R}^2$, has constant $y$ value if and only if the orthogonal complement of the normal vector of $P$ has zero $y$-component, or equivalently, the normal vector is pointing in the $y$-direction.} \label{fig:cts7}
\end{figure}

\begin{Lemma} \label{lem:fvv}
  For $V' \subset V$, $v \in {V'}\setcomp$ is $\Lambda_0$-dependent on $V'$ if and only if for each $f \in \BWset$, $f_v = \sum_{v'\in V'}c_{v'}f_{v'}$ with coefficients $c_{v'}$ independent of $f$.
\end{Lemma}
We do not require that the linear combination in the statement of the lemma is unique.
\begin{IEEEproof}
  From the definition of $\Lambda_0$ in \cref{Lambda0}, for each $t\in\Real$, we have $\bU_{\Lambda_0,V}\mbf_V(t)=0$. Therefore, 
  \begin{align*}
    [\bU_{\Lambda_0,V'},\, \bU_{\Lambda_0,{V'}\setcomp}][\mbf_{V'}(t);\, \mbf_{{V'}\setcomp}(t)] = 0
  \end{align*}
  Geometrically, any $f$ with fixed components $f_{V'} = (f_{v'})_{v'\in V'}$ can be viewed as the intersection of $|\Lambda_0|$ hyperplanes, with normal vectors the rows of $\bU_{\Lambda_0,{V'}\setcomp}$. For $v\in V\backslash V'$, the intersection has a constant $v$-component, i.e., determined uniquely by $f_{V'}$ as a linear combination of its components, if and only if each of the normal vector is is parallel to the axis of $v$. The latter condition is equivalent to: the orthogonal complement of these normal vectors has zero $v$-component (as illustrated in \cref{fig:cts7}).
\end{IEEEproof}

The geometric condition given in \cref{defn:lvba} allows one to directly check $\Lambda_0$-dependence of $v$ on $V'$. However, in our theoretical study, we mainly use the equivalent condition given in \cref{lem:fvv}.

The notion of ``dependency'' leads to a \emph{matroid} structure of $V$, to which we give a self-contained exposition in \cref{sec:mat}. Using the theory of matroid, we can apply the greedy algorithm \cite{Oxl92} if we want to optimize a function on $V$.

\begin{Proposition} \label{thm:iuit}
  Consider a space of continuous-time graph signals $\BWset$.
  \begin{enumerate}[(a)]
    \item \label{it:iui} $\BWset$ is tight if and only if for any $V'\subset V$ and vertex $v\in {V'}\setcomp$ that is $\Lambda_0$-dependent on $V'$, we have
          \begin{align}
            \calB[v]\leq \max_{v'\in V'}\calB[v'].
          \end{align}
          In particular, if $\BWset$ is tight, then for any $V_0 \in \calU(\Lambda_0)$, we have
          \begin{align}\label{V0tight}
            \max_{v\in V} \calB[v] = \max_{v\in V_0} \calB[v].
          \end{align}

    \item If $\BWset$ is tight, then there is an $f \in \BWset$ such that $f_v$ has bandwidth exactly $\calB[v]$ for all $v\in V$.
  \end{enumerate}
\end{Proposition}
\begin{IEEEproof}
  \begin{enumerate}[(a)]
    \item Suppose $\BWset$ is tight, $V' \in V$ and $v$ is $\Lambda_0$-dependent on $V'$. For any $f\in \BWset$, by \cref{lem:fvv}, $f_v$ is a linear combination of $\set{f_{v'} \given v'\in V'}$. Therefore, the bandwidth of $f_v$ is bounded by $\max_{v'\in V'}\calB[v']$. By tightness, we have $\calB[v]\leq \max_{v'\in V'}\calB[v']$.

          Conversely, suppose $\calB[v]\leq \max_{v'\in V'} \calB[v']$ as long as $v$ is $\Lambda_0$-dependent on $V'$. Choose $V_0$ in $\calU(\Lambda_0)$ that minimizes $\sum_{v'\in V_0}\calB[v']$. We index $v'\in V_0$ according to increasing order of $\calB[v']$ to obtain $V_0=\set{v_1,v_2,\ldots,v_{|V_0|}}$. For an arbitrary $v\in V_0\setcomp$, let $k$ be the smallest index such that $v$ is $\Lambda_0$-dependent on $\{v_1,\ldots, v_k\}$. By our assumption, we have $\calB[v]\leq \calB[v_k]$. Moreover, as $v$ is not $\Lambda_0$-dependent on $\{v_1,\ldots, v_{k-1}\}$, $v_k$ is $\Lambda_0$-dependent on $\{v_1,\ldots, v_{k-1},v\}$ by \cref{lem:fvv}. This implies that $V_0\cup \{v\}\backslash \{v_k\} \in \calU(\Lambda_0)$. By the minimality in choosing $V_0$, we have $\calB[v] = \calB[v_k]$.

          We can choose $f\in\BWset$ such that $f_{v_k}$ has bandwidth exactly $\calB[v]$ and $f_{v_i} =0$ for all $v_i\neq v_k$ in $V_0$. Since $f_v$ is a linear combination of $\set{f_{v_i} \given i \leq k}$, this construction allows us to choose $f_v=f_{v_k}$ having bandwidth exactly $\calB[v]$. Repeating the same argument for every vertex, we have shown that $\BWset$ is tight. Finally, by considering $V'=V_0\in\calU(\Lambda_0)$, we obtain \cref{V0tight}.

    \item Choose $V_0\in \calU(\Lambda_0)$ that minimizes $\sum_{v'\in V_0}\calB[v']$. Same as in part \ref{it:iui}, we index the vertices in $V_0$ in increasing order of $\calB[v']$ to obtain $V_0=\set{v_1,v_2,\ldots,v_{|V_0|}}$. As earlier, for any $v\notin V_0$, we find the smallest $k\leq |V_0|$ such that $v$ is $\Lambda_0$-dependent on $\{v_1,\ldots, v_k\}$. As shown in part \ref{it:iui}, $\calB[v]=\calB[v_k]$. By tightness, we choose $f$ such that the bandwidth of $f_{v_j}$ is $\calB[v_j]$, for all $j\leq k$. The signal $f_v$ is a linear combination of $\set{f_{v_j}\given j\leq k}$. We can always change $f_{v_k}$ by a nonzero scalar multiple to $y f_{v_k}$ to ensure $f_{v_k}$ is not canceled out in the linear combination that makes up $f_v$. This a possibility since $v$ is not $\Lambda_0$-dependent on $\{v_1,\ldots, v_{k-1}\}$. Consequently, $f_v$ has bandwidth exactly $\calB[v_k]$. All but finitely many such $y$ makes this hold.

    This procedure can be repeated for all the vertices of $V$. Each time we may perform a scaling of $f_{v_j}$ for some $j\leq |V_0|$. However, from the previous paragraph, we are restricted from choosing the scaling coefficient from a finite set. There is always a set of scaling coefficients $y_k$ of $f_{v_k}$,$1\leq k\leq |V_0|$ such that the resulting $f_v$ has bandwidth exactly $\calB[v]$ for all $v\in V$.
  \end{enumerate}
\end{IEEEproof}

We now state the main sampling result for $\BWset$. We first define a partial order on collections of bandwidths: $\calB \leq \calB'$ if $\calB[v] \leq \calB'[v]$ for each $v\in V$. The first part states that $\BWset$ can be replaced by a maximal tight one, and the second part is the generalized Nyquist-Shannon sampling theorem with an explicit formula for the minimal sampling rate.

\begin{Theorem} \label{thm:feu}
  \begin{enumerate}[(a)]
    \item\label{it:few} For every $\BWset$, there is a unique maximal $\calB^*\leq \calB$ such that $W_{\calB^*,\calC}$ is tight. Moreover, $\BWset = W_{\calB^*, \calC}$.

    \item\label{it:GNSST} (Generalized Nyquist-Shannon sampling theorem) For every $\BWset$ with simple GFT bandwidth, the minimal sampling rate is
          \begin{align}
            r^* = \min_{V_0\in \calU(\Lambda_0)} 2\sum_{v\in V_0}\calB[v].
          \end{align}

          Moreover, there exists discrete sampling set $S$ with sample rate $r^*$ such that $S$ determines a unique signal $f \in \BWset$.
  \end{enumerate}
\end{Theorem}

Any $V_0$ giving rise to $r^*$ is called a \emph{minimal vertex set} \gls{wrt}\ $\Lambda_0$.


\begin{IEEEproof}
\begin{enumerate}[(a)]
\item The proof of this part is not used in the sequel and is provided in \cref{sec:tig}.
\item To show the optimal sample rate is $\min_{V_0\in \calU(\Lambda_0)} 2\sum_{v\in V_0}\calB[v]$, we want to invoke Theorem 1 or (19) of \cite{Ven04}. Let $V_0$ be a minimal vertex set. We first notice that any signal $f$ is uniquely determined by $f_v, v\in V_0$ as $V_0\in \calU(\Lambda_0)$. We want to show that given $f_v$ bandlimited by $\calB[v]$ for each $v\in V_0$, the corresponding $f$ indeed belongs to $\BWset$.

  Consider any $v'\notin V_0$. As earlier, we may index $V_0 =\{v_1,\ldots,v_{|V_0|}\}$ according to increasing order of $b_{v_i}, 1\leq i\leq |V_0|$. If $v'$ is $\Lambda_0$-dependent on $v_1,\ldots, v_k$ such that $k$ is the smallest, then $\calB[v']\geq \calB[v_k]$. However, by \cref{lem:fvv}, $f_{v'}$ is the linear combination of $f_{v_1},\ldots,f_{v_k}$. Hence, the bandwidth of $f_{v'}$ is bounded by $\calB[v_k]\leq \calB[v']$. Therefore, $f\in \BWset$.

  Now, along each $v\in V_0$, we choose a discrete sampling set $S_v$ uniformly at the Nyquist-Shannon rate $2\calB[v]$. The corresponding base functions are translates of the $\sinc$ function along $v$ and $0$ elsewhere. We have shown that any $f\in \BWset$ is uniquely determined by $f_v, v\in V_0$. In turn, each $f_v$ is uniquely determined by their value at $S_v$ by the Nyquist-Shannon theorem (as Hilbert space isomorphism). The domain of the isomorphism can be viewed as $|V_0|$ channels indexed by $v\in V_0$, with the bandwidth of the $v$-th channel $\calB[v]$. Therefore, the rate $r^*$ is necessary by Theorem 1 of \cite{Ven04}.
\end{enumerate}
\end{IEEEproof}

For each $v$, there is an associated $\calB[v]\geq 0$, which can be viewed as a function on the matroid $V$. Therefore, to minimize $2\sum_{v\in V_0}\calB[v]$ over all $V_0$, it suffices to apply the greedy algorithm \cite{Oxl92}: Starting from the empty set $V_0=\emptyset$, in each iteration, we add $v$ to $V_0$ such that $\calB[v]$ is the smallest among all $v$'s that are not $\Lambda_0$-dependent on $V_0$. On the other hand, a tight $\calB^*$ has essentially gotten rid of all redundant information.

\section{Admissible sequence of vertices and sampling} \label{sec:adm}
In \cref{sec:red}, we describe how to obtain a filtration $W_0 \subset W_1 \subset \ldots \subset W_k = \BWset$ such that each subquotient $W_{i}/W_{i-1}$ is a bandlimited subspace of $L^2(\mathbb{R})$ (cf.\ \cref{prop:tio}). The fundamental question we want to address is the following:

\emph{Given a surjective Hilbert space morphism: $\alpha: H_1 \to H_2$, if we know how to sample for $H_2$ and $\ker(\alpha)$, how can one sample for $H_1$?}

\begin{figure}[!htb]
  \centering
  \includegraphics[width=0.9\columnwidth]{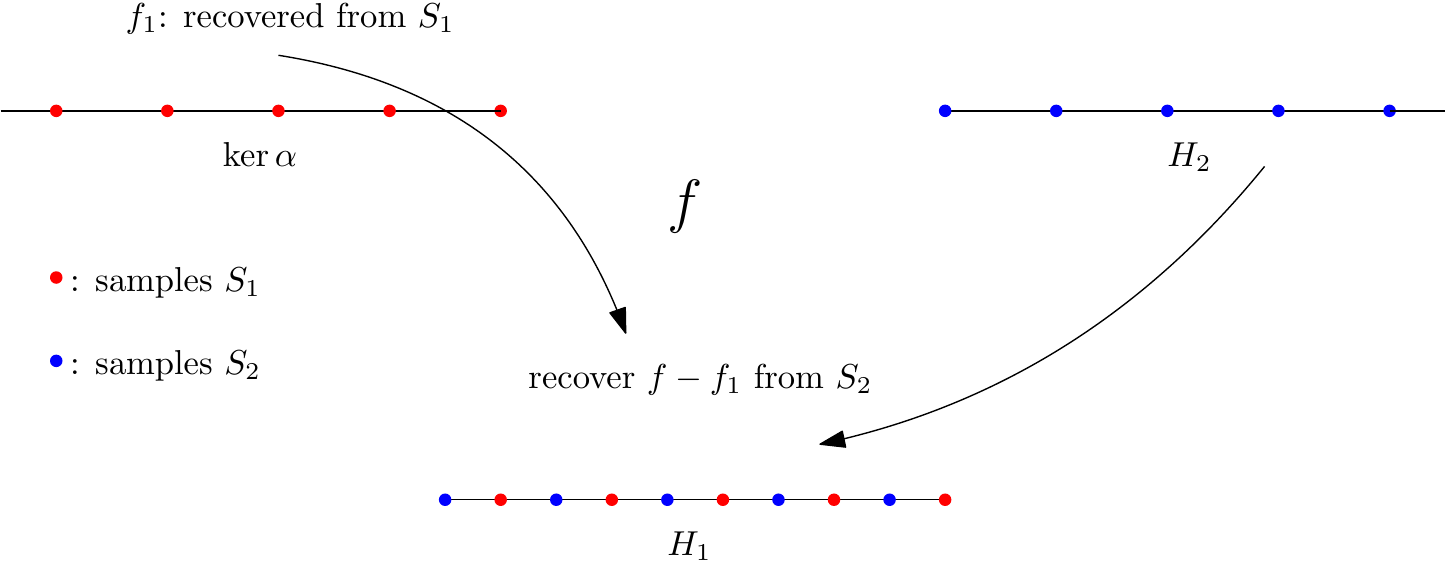}
  \caption{Illustration of the overall recovering scheme.} \label{fig:cts3}
\end{figure}

The idea is as follows. Suppose $f \in H_1$ is given. To solve the sampling problem, we find a sample set $S_1$ such that we can uniquely recover $f_1\in\ker(\alpha)$ with observations made at $S_1$. Assuming we know how to sample for $H_2$, we find such a sample set $S_2$. We first update the observations at $S_2$ by subtracting the values of $f_1$ at $S_2$ to obtain samples of $f_2 = f - f_1$ at $S_2$ together with the requirement that $f_2$ vanishes on $S_1$. We can then recover $f_2$ and finally obtain $f = f_1 + f_2$. The scheme is illustrated in \cref{fig:cts3}.

Now we apply this idea to $W_0 \subset W_1 \subset \ldots \subset W_k = \BWset$. The filtration and sampling for the subquotients have the following properties:
\begin{enumerate}[P.1]
  \item\label{P1} There is a filtration of subsets of $\Lambda$: For each $i=0,\ldots,k$, let $W_i=W_{\calB,\calC_i}$ and $\Lambda_{i,0}=\Lambda_0(\calC_i)=\set{\lambda\in\Lambda\given \calC_i[\lambda]=0}$. Then $\Lambda_{0,0} \supset \Lambda_{1,0} \ldots \supset \Lambda_{k,0}$.
  \item\label{P2} Each $W_{i}/W_{i-1}$, $1\leq i\leq k$, can be identified with a bandlimited subspace of $L^2(\mathbb{R})$. Let $b_i$ be its bandwidth.
\end{enumerate}

From \cref{prop:tio} and the discussion that follows, to sample for each $W_{i}/W_{i-1}$, $1\leq i\leq k$, we carry out the following steps:
\begin{enumerate}[(i)]
  \item Find a uniqueness set $V_{i}$ \gls{wrt} $\Lambda_{i,0}$ (i.e., $\bU_{\Lambda_{i,0},V_{i}\setcomp}$ is invertible; see \cref{defn:fps}), and $v_{i} \in V_{i}$ such that:
        \begin{enumerate}[(a)]
          \item $\bx_{V_{i}}[v_{i}] \ne 0$ (cf.\ \cref{bxV});
          \item $\calB[v_{i}]\geq b_{i}$; and
          \item $\calB[v] \geq b_i$ for $v \notin V_{i}$ that is not $\Lambda_{i,0}$-dependent on $V_{i-1}$.
        \end{enumerate}
  \item We sample along $v_{i}$ with rate $2b_i$ and set the vertex signal to be $0$ at $v\in V_{i}\setcomp$.
\end{enumerate}

We now need to find a way to put together the individual sampling method for each layer of the filtration. For this, we introduce the following notion.

\begin{Definition} \label{defn:asv}
  Continuing from \ref{P1}, a sequence $V_0 \subset V_1 \ldots \subset V_k$ of subset of vertices is called an \emph{admissible sequence} if the following holds:
  \begin{enumerate}[(a)]
    \item\label{asv:1} $V_0$ is a minimal vertex set \gls{wrt} $\Lambda_{0,0}$ (cf.\ \cref{thm:feu}).
    \item\label{asv:2} For each $i=1,\ldots,k$, $V_i$ is a uniqueness set \gls{wrt} $\Lambda_{i,0}$.
    \item\label{asv:3} $V_i\backslash V_{i-1}$ is a singleton $\{v_i\}$.
    \item\label{asv:4} $\bx_{V_i}[v_i] \ne 0$, $\calB[v_i]\geq b_i$ and $\calB[v] \geq b_i$ for all $v\notin V_i$ that is not $\Lambda_{i,0}$-dependent on $V_{i-1}$.
  \end{enumerate}
\end{Definition}

\begin{Theorem}
  Using the definitions in \ref{P1} and \ref{P2}, if there is an admissible sequence of vertex sets $V_0\subset V_1\subset \ldots \subset V_k$, then the minimal sample rate for perfect recovery of signals in $\BWset$ is 
  \begin{align}
  2\parens*{\sum_{v\in V_0}\calB[v]+\sum_{1\leq i\leq k}b_i}.
  \end{align}
\end{Theorem}
\begin{IEEEproof}
  We can sample at the sample set $S_0$ (a discrete subset of $V_0\times\Real$; see \cref{defn:lvba}) along $V_0$ with rate $2\sum_{v\in V_0}\calB[v]$, and sample at the sample set $S_i$ along $v_i$, $i\geq 1$ with rate $2b_i$ for each $1\leq i\leq k$. The overall sampling rate is thus $2(\sum_{v\in V_0}\calB[v]+\sum_{1\leq i\leq k}b_i)$.

  For recovery, we proceed inductively. From \cref{thm:feu}\ref{it:GNSST}, since $W_0$ has simple GFT bandwidths, we first reconstruct an $f_0 \in W_0$ with observations from $S_0$, and update the observations at the remaining sample sets $S_1, \ldots, S_k$ by subtracting $f_0(s)$ at each $s\in S_i$, $1\leq i\leq k$. We recover $f_1$ with observations at $S_1$ and $0$ at $S_0$. We can do this by conditions \ref{asv:2}--\ref{asv:4} of \cref{defn:asv}. We follow the steps \ref{it:rau}--\ref{it:avc} in \cref{sec:str} (with $v_1$ here as $v$ and $V_1$ here as $V_0$ in those steps). The observations at $S_2\cup \ldots \cup S_k$ are updated accordingly. We proceed similarly to recover $f_i, 2\leq i\leq k$ with (updated) observations at $S_i$ and $0$ at $S_0\cup \ldots \cup S_{i-1}$. The signal we want to recover is nothing but $f_1 + \ldots + f_k$.

  We use translates for the $\sinc$ functions to construct $f_1,\ldots, f_k$. The stated rate is the minimal sample rate by Theorem 1 of \cite{Ven04}.
\end{IEEEproof}

This result can be viewed as the Nyquist-Shannon theorem for general $\BWset$. There are various sets of conditions that ensure the existence of an admissible sequence of sets of vertices, we provide one such case as follows.

\begin{Lemma}
  Suppose the following holds:
  \begin{enumerate}[(1)]
    \item \label{it:tiam} There is a minimal vertices set $V_0$ such that $\max_{v\in V_0} \calB[v] \leq \min_{1\leq i\leq k} b_i$.
    \item \label{it:fev} If $V$ and $V\cup \{v\}$ are uniqueness sets \gls{wrt}\ $\Lambda_i$ and $\Lambda_{i+1}$ respectively, then $\bx_{V_{i+1}}[v] \neq 0$.
  \end{enumerate}
  Then there is an admissible sequence of sets of vertices.
\end{Lemma}

\begin{IEEEproof}
  We choose $V_0$ as in \ref{it:tiam}. Suppose we have chosen $V_{i-1}$. We find $v_i$ such that
  \begin{enumerate}[(a)]
    \item \label{it:vvc} $V_{i} = V_{i-1}\cup \{v_i\}$ is a uniqueness set and hence $\bx_{V_{i+1}}[v]\neq 0$ by \ref{it:fev}; and
    \item \label{it:bit}$\calB[v_i]$ is the smallest among all $v_i$ satisfying \ref{it:vvc}.
  \end{enumerate}
  We claim that the procedure yields an admissible sequence of sets of vertices. As $V_i$ is a uniqueness set and  $\max_{v\in V_0} \calB[v] \leq \min_{1\leq i\leq k} b_i$, we have $\calB[v_i]\geq b_i$. It remains to show that $\calB[v] \geq b_i$ for $v\notin V_i$ that does not depend on $V_{i-1}$.

  Suppose there is a $v\notin V_i$ that does not depend on $V_{i-1}$ and $\calB[v]<b_i \leq \calB[v_i]$. We may express $f_v$ as a linear combination of $f_{v'}, v'\in V_{i-1}$ and $f_{v_i}$, with nonzero coefficient for $f_{v_i}$. As a consequence, $v_i$ depends on $V_{i-1}\cup \{v\}$. Therefore, $V_{i-1}\cup \{v\}$ is also a uniqueness set. This contradicts the minimality condition~\ref{it:bit} in choosing $v_i$.
\end{IEEEproof}

We now describe yet another modification to the above sampling scheme. For the above sampling scheme, we have the simplified summary: we have two vertices $v_1, v_2$ and samples $S_1$ and $S_2$ along $v_1$ and $v_2$ respectively. Assume the sample rates for $S_1, S_2$ are $2B, 2C$ respectively and $B\leq C$. We propose to recover $f_1$ using observations at $S_1$, and $f_2$ using observations at $S_2$ and assuming $f_2$ is $0$ at $S_1$.

\begin{figure}[!htb]
  \centering
  \includegraphics[width=0.8\columnwidth]{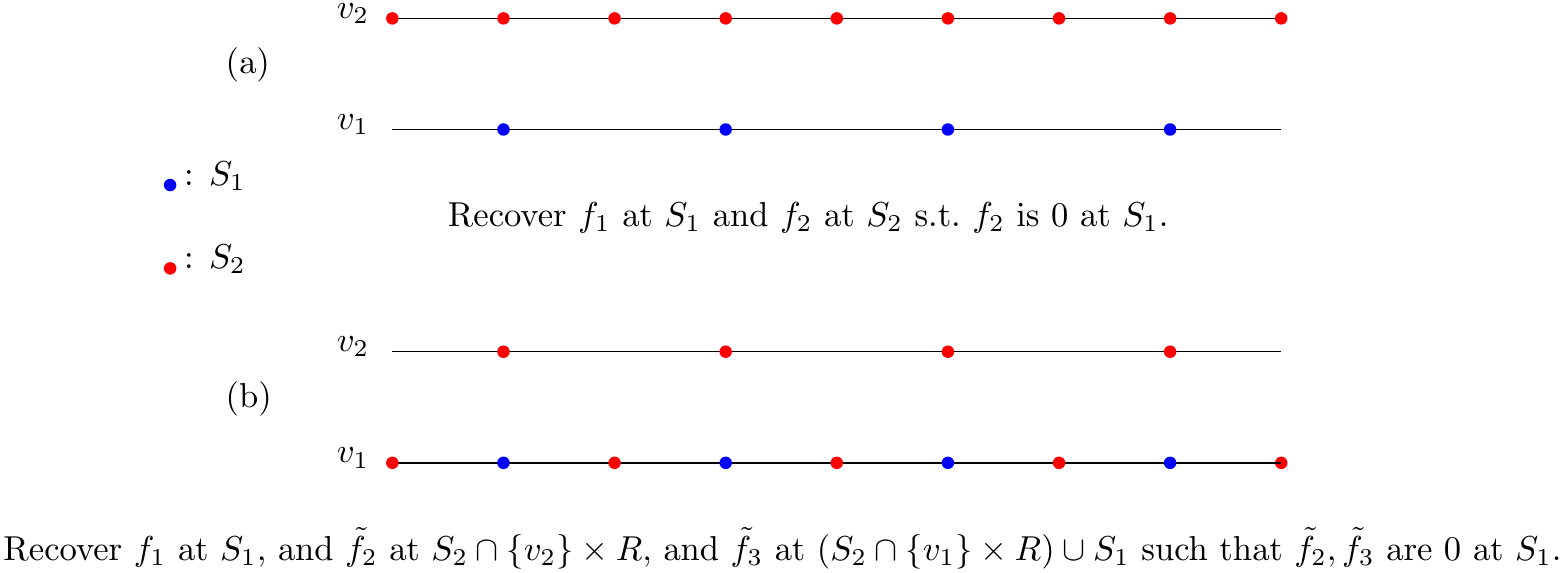}
  \caption{Illustration of the alternative sampling scheme.} \label{fig:cts6}.
\end{figure}

However, we may encounter situations that at $v_2$, we are not allowed to have a signal whose bandwidth is $C$. This means we cannot sample at a rate $2C$ along $v_2$. On the other hand, if we can sample at rate $2B$ at $v_2$ and $2C$ at $v_1$, we may have an alternate sampling scheme (illustrated in \cref{fig:cts6}). We keep the same $S_1$, and $S_2$ consists of a set with rate $2B$ along $v_2$ and a set of rate $2(C-B)$ along $v_1$. For recovery, we first get the same $f_1$ using $S_1$. Then we recover $\tilde{f}_2$ with $S_2\cap \{v_2\}\times R$, and $\tilde{f}_3$ with $(S_2\cap \{v_1\}\times R) \cup S_1$ such that $\tilde{f}_2, \tilde{f}_3$ are $0$ at $S_1$. The signal to be recovered is $f_1+\tilde{f}_2+\tilde{f}_3$.

If we view each $v\in V$ as a ``sensor'' taking measurements, then the sampling scheme discussed in this section concentrates on a few sensors. On the other hand, it is possible that there are spare sensors; and if we make measurements at the spare sensors, we may reduce the sample rate at each sensor. This leads to the notion of ``eccentricity'' of sampling, which is discussed in \cref{sec:ecc}.

\section{An explicit example}\label{sec:eg}

\begin{figure}[!htb]
  \centering
  \includegraphics[width=0.3\columnwidth]{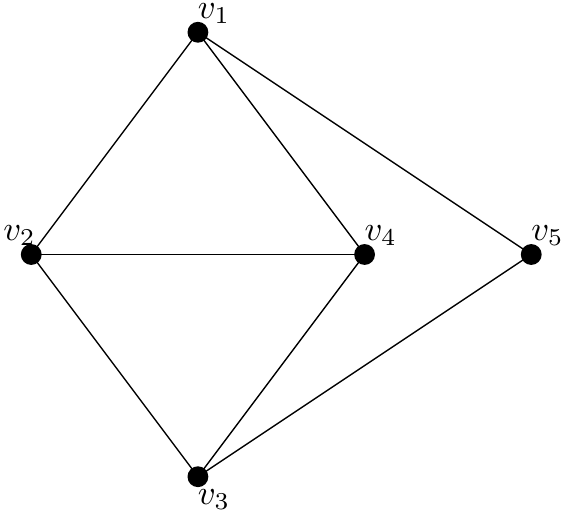}
  \caption{The graph $G$ with $5$ vertices and $7$ edges.} \label{fig:cts2}
\end{figure}

In this section, we work out an explicit example to demonstrate the entire sampling scheme. Suppose the graph $G=(V,E)$ has $5$ vertices and is shown in Figure~\ref{fig:cts2}. Its Laplacian matrix $L$ is
\begin{align*}
  L = \begin{bmatrix}
    3  & -1 & 0  & -1 & -1 \\
    -1 & 3  & -1 & -1 & 0  \\
    0  & -1 & 3  & -1 & -1 \\
    -1 & -1 & -1 & 3  & 0  \\
    -1 & 0  & -1 & 0  & 2
  \end{bmatrix}.
\end{align*}

The frequencies are $\Lambda = \{\lambda_1, \lambda_2, \lambda_3, \lambda_4, \lambda_5 \}$ associated with eigenvalues $0,2,3,4,5$ respectively. We randomly generate $\calB = \{\calB[v_1]=5, \calB[v_2]=5, \calB[v_3]=1, \calB[v_4]=4, \calB[v_5]=4\}$ and $\calC= \{\calC[\lambda_1]=9, \calC[\lambda_2]=2, \calC[\lambda_3]=5, \calC[\lambda_4]=\infty, \calC[\lambda_5]=\infty\}$.

We first follow \cref{sec:red} to find a filtration: $W_0 \subset W_1 \subset W_2 \subset \BWset$ such that $W_0$ has simple GFT bandwidth. The intermediate sub-quotients are discussed as follows.

\begin{itemize}
  \item $\BWset/W_2$: We identify $\lambda^* = \lambda_2$ and $\Lambda_0=\emptyset$. The only choice of $V_0$ is $V$. As the eigenvector associated with $\lambda_2 $ is $(0,0.408,0,0.408,-0.817)^T$, it is straightforward to find that $b_{\lambda^*} = 2$. Therefore, $W_2= W_{\calB, \calC_2}$ where $\calC_2=\{\calC_2[\lambda_1]=9, \calC_2[\lambda_2]=0, \calC_2[\lambda_3]=5, \calC_2[\lambda_4]=\infty, \calC_2[\lambda_5]=\infty\}$. Moreover, the bandwidth of $\BWset/W_2$ is exactly $2$.

  \item $W_2/W_1$: To pass from $W_2$ to $W_1$, $\lambda^* = \lambda_3$, $\Lambda_0 = \lambda_2$ and $$\calU(\Lambda_0) = \{V\backslash\{v_2\},V\backslash\{v_4\},V\backslash\{v_5\}\}.$$ Any $V_0 \in \calU(\Lambda_0)$ contains $v_1$ and $\bx_{V_0}[v_1]\neq 0, \calB[v_1]=5$. Therefore, $b_{V_0}\geq 5$ and $b_{\lambda^*}=5$. Hence, $W_1 = W_{\calB,\calC_1}$ where $\calC_1=\{\calC_1[\lambda_1]=9, \calC_1[\lambda_2]=0, \calC_1[\lambda_3]=0, \calC_1[\lambda_4]=\infty, \calC_1[\lambda_5]=\infty\}$. The bandwidth of $W_2/W_1$ is exactly $5$.

  \item $W_1/W_0$: Finally, $\lambda^*=\lambda_1$ and $\Lambda_0 = \{\lambda_2,\lambda_3\}$. As $\calC_1[\lambda^*] = 9 > \max_{v\in V} \calB[v]$, we have $b_{\lambda^*} = b_{V_0}$ for some $V_0\in \calU(\Lambda_0)$. One may find that $\bx_{V_0}[v]\neq 0$ for every $v\in V, V_0\in \calU(\Lambda_0)$. As any $V_0\in\calU$ contains at least three vertices, hence $b_{V_0}=5$ and so is $b_{\lambda^*}$. Therefore, $W_0 = W_{\calB,\calC_0}$ where $\calC_0=\{\calC_0[\lambda_1]=0, \calC_0[\lambda_2]=0, \calC_0[\lambda_3]=0, \calC_0[\lambda_4]=\infty, \calC_0[\lambda_5]=\infty\}$. The bandwidth of $W_1/W_0$ is exactly $5$.
\end{itemize}

For the final step, we handle $W_0$ using \cref{sec:simple}. It has simple GFT bandwidth and $\Lambda_0 = \{\lambda_1,\lambda_2,\lambda_3\}$. We notice that $\{v_3,v_5\} \notin \calU(\Lambda_0)$. The other choice is $\{v_3,v_4\} \in \calU(\Lambda_0)$. By Theorem~\ref{thm:feu}, the minimal sample rate for perfect recovery of signals in $W_0$ is $2(4+1) = 10$. We may further verify that $v_5$ depends on $\{v_3\}$. As $b_{v_5}>b_{v_3}$, $\calB$ is not tight \gls{wrt}\ $B'_2$. We can sample at rate $2$ along $v_3$ and at rate $8$ along $v_4$. The eccentricity (c.f.\ Appendix~\ref{sec:ecc}) for this sampling scheme is $4$. Moreover, by invoking Proposition~\ref{prop:svo}, we can also sample at rate $2$ along $v_3$, at rate $4$ along $v_2, v_4$. The eccentricity (c.f.\ Appendix~\ref{sec:ecc}) of this sampling scheme is only $2$.

We have an admissible sequence of sets of vertices: 
$\{v_3,v_4\} \subset \{v_3,v_4,v_5\} \subset \{v_1,v_3,v_4,v_5\} \subset \{v_1,v_2,v_3,v_4,v_5\}$. As a consequence, the overall minimal sample rate is $34$.

\section{Conclusions} \label{sec:con}

In this paper, we give a complete description of the sampling theory for bandlimited continuous-time graph signal. The highlight is the generalization of the celebrated Nyquist-Shannon sample theorem. The latter has an enormous amount of applications, and for future work, we shall focus on applying our results to real network data analysis problems.

\appendices

\section{Matroid Characterization} \label[Appendix]{sec:mat}

In \cref{sec:simple}, we formally introduced the notion of dependence on $V$. Correspondingly, we can define a set of vertices being independent. This falls within the general framework of matroid theory, which is convenient for studying such a property.

Recall that a \emph{finite matroid} \cite{Oxl92} is a pair $(E,\mathcal{I})$, where $E$ is a finite set and $\mathcal{I}$ is a family of subsets of $E$ (called \emph{independence sets}) with the following properties:
\begin{enumerate}[(1)]
  \item The empty set is independent.
  \item Every subset of independent set is independent.
        \item\label{it:iaab} If $A$ and $B$ are two independent sets and $A$ has more elements than $B$, then there exists $x\in A\backslash B$ such that $B\cup \{x\}$ is in $\mathcal{I}$.
\end{enumerate}

Based on \cref{defn:lvba} of \cref{sec:simple}, we define a set of vertices $V'\subset V$ to be an \emph{independent set} of vertices if $v'$ does not depend on $V'\backslash \{v'\}$ for any $v'\in V'$. It is worth pointing out that this notion of independence relies on the choice $\Lambda_0$, as so does \cref{defn:lvba}.

\begin{Proposition}
  $V$ with the above notion of independence is a matroid.
\end{Proposition}

\begin{IEEEproof}
  The first two conditions of a matroid follow directly from the definition. Let us verify Condition~\ref{it:iaab} of a matroid.

  Suppose $V_1$ and $V_2$ are independent and $|V_1|=|V_2|+1$. We order the elements of $V_1$ as $v_1,\ldots, v_{k+1}$; and order the elements of $V_2$ as $u_1,\ldots, u_k$. For $v_1$, if it is independent of $V_2$, then we are done. Otherwise, by \cref{lem:fvv}, for any signal $f$, $f_{v_1}$ is a linear combination of $f_{v_1} = \sum_{1\leq j\leq k}c_jf_{u_j}$, with coefficients $c_j$ independent of $f$. Some $c_j$, say $c_1$, is non-zero. Hence, $u_1$ depends on $V' = \{v_1, u_2, \ldots, u_k\}$. Moreover, any $v$ depends on $V_2$ if and only if it depends on $V'$.

  For $v_2$, if it is independent of $V'$, then again we are done. Otherwise, for any $f$, $f_{v_2}$ is a linear combination $f_{v_2} = d_1f_{v_1} + \sum_{2\leq j\leq k} c_jf_{u_j}$ ($c_j$'s are in general different from those in the previous paragraph). As $v_2$ is independent of $v_1$, some $c_j$, say $c_2$ is non-zero. Hence, $u_2$ depends on $V' = \{v_1,v_2, u_3,\ldots, u_k\}$. Moreover, any $v$ depends on $V_2$ if and only if it depends on $V'$.

  We proceed with $v_3, \ldots, v_k$ in the same way. If none of them is independent of $V_2$, by the same argument, we have any $v$ depends on $V_2$ if and only if it depends on $V'=\{v_1, \ldots, v_k\}$. However, $v_{k+1}$ is independent of $V'$ and hence $v_{k+1}$ is independent of $V_2$.
\end{IEEEproof}

A maximal independent set of a matroid is called a basis. By Condition~\ref{it:iaab}, all bases have the same size. One important aspect of the matroid theory is that to choose a basis that minimizes a function on the matroid, one can simply apply the greedy algorithm \cite{Sid18}.

\section{Tightness} \label[Appendix]{sec:tig}

In this appendix, we supply further discussion on tightness and prove the first half of \cref{thm:feu}. This is a supplement of \cref{sec:simple}; and we retain the notations of \cref{sec:simple} such as $\Lambda_0$, $\calU(\Lambda_0)$.

We first define how to take union of sets of bandwidths. For $\calB=\{\calB[v],v\in V\}$ and $\calB'=\{\calB'[v],v\in V\}$, we may take their union $\calB\cup \calB' = \{\calB\cup \calB'[v], v\in V\}$, where $\calB\cup \calB'[v] = \max \{\calB[v],\calB'[v]\}$. Therefore, $\calB \leq \calB'$ if and only if $\calB\cup \calB' = \calB'$.

\begin{Lemma} \label{lem:ibb}
  If both $\calB$ and $\calB'$ are tight \gls{wrt}\ $\calC$, then so is $\calB\cup \calB'$.
\end{Lemma}

\begin{IEEEproof}
For $v\in V$, without loss of generality, we assume that $\calB\cup \calB'[v]= \calB[v]$. By tightness of $\calB$, there is an $f$ whose bandwidth at $v$ is exactly $\calB[v]$. At any other vertex $u\in V$, the bandwidth of $f$ is bounded by $\calB[u]$, which is in turn bounded by $\calB\cup \calB'[u]$. Therefore, $\calB\cup \calB'$ is tight \gls{wrt}\ $\calC$ by definition.
\end{IEEEproof}

\begin{Lemma} \label{lem:fau}
  For any $\BWset$, suppose $\calB'\leq \calB$ is tight \gls{wrt}\ $\calC$. Then there is a $V_0\in \calU(\Lambda_0)$ and $\calB' \leq \calB_{V_0}=\{\calB_{V_0}[v], v\in V\} \leq \calB$ such that
  \begin{enumerate}[(1)]
    \item $W_{\calB_{V_0}, \calC}$ is tight.
    \item $\calB_{V_0}[v] = \calB[v]$ for every $v\in V_0$.
  \end{enumerate}
  Moreover, there is a unique maximal $\calB_{V_0}$ such that the above holds.
\end{Lemma}

\begin{IEEEproof}
  Suppose there is a $V_0\in \calU(\Lambda_0)$ such that $\calB'[v] = \calB[v]$ for every $v\in V_0$. Among all such $V_0$, we may choose one such that $\sum_{v\in V_0}\calB[v]$ is minimized. We order the vertices of $V_0=\{v_1,\ldots, v_m\}$ increasingly according to $\calB[v], v\in V_0$. If $v\in V$ depends on $\{v_1,\ldots, v_k\}$ such that $k$ is the smallest, as in the proof of Proposition~\ref{thm:iuit} using minimality of $\sum_{v\in V_0}\calB[v]$, we have $\calB[v]= \calB[v_k]$. If we define $\calB_{V_0}[v]=\calB[v_k]$ as such, $\calB_{V_0}$ is tight and maximal. On the other hand, by tightness of $\calB'$, $\calB'[v] \leq \calB[v]=\calB[v_k]$ and hence $\calB'\leq \calB_{V_0}$.

  Now, suppose that there is no $V_0 \in \calU$ such that ``$\calB'[v] = \calB[v]$ for every $v\in V_0$'' holds. We want to proceed inductively. Let $V_0'=\{v_1,\ldots, v_m\}$ be the maximal independent subset of vertices such that $\calB'[v] = \calB[v], v\in V_0'$. Let $v$ be the vertices such that $\calB[v]$ is the smallest among those independent of $V_0'$. By our assumption, $\calB[v]> \calB'[v]$. We may enlarge $V_0'$ by replacing it with $\{v_1,\ldots,v_m\} \cup \{v\}$, also denoted by $V_0'$.

  To introduce $\calB'_{V_0'}$, for each $v'\in V$, we can define $\calB'_{V_0'}[v'] = \max\{\calB[v], \calB'[v]\}$ for $v'$ depends on (the enlarged new) $V_0'$ and independent of $\{v_1,\ldots, v_m\}$; and retain $\calB'[v']$ for the remaining vertices. The minimality of $\calB'[v]$ when we choose $v$ ensures that $\calB'\leq \calB'_{V_0'} \leq \calB$. However, we have enlarged maximal independent subset of vertices with bandwidth $\calB[v]$. To conclude by induction, it suffices to verify that $\calB'_{V_0'}$ is tight. To see this, choose any $V_0\in \calU(\Lambda_0)$ containing $V_0'$, we let $f$ have bandwidth exactly $\calB[v]$ at $v$ and $0$ at the remaining vertices of $V_0$. Thus, the bandwidth of $f_{v'}$ at the vertex $v'$, where a change has been made, is exactly $\calB[v]=\calB_{V'_0}[v']$. Combined with the tightness of $\calB'$, for any $v'\in V$, we can find an appropriate $f$ with bandwidth $\calB_{V_0'}[v']$. Hence, $\calB'_{V_0'}$ is tight.
\end{IEEEproof}

\begin{IEEEproof}
  (\cref{thm:feu}\ref{it:few}) By \cref{lem:fau}, there is a finite set $\scB$ of $\calB_{V_0} \subset \calB$ tight \gls{wrt}\ $\calC$ such that: each $\calB'\leq \calB$ tight \gls{wrt}\ $\calC$ is contained in some $\calB_{V_0}\in \scB$. By \cref{lem:ibb}, $\calB^*$ can be constructed by taking the union $\cup_{B_{V_0}\in \scB}B_{V_0}$.

  As $\calB^{*} \leq \calB$, hence $W_{\calB^*,\calC}\subset W_{\calB, \calC}$. Conversely, suppose $f\in \BWset$. Let $\calB'$ be defined such that $\calB'[v]$ is the bandwidth of $f_v$ at $v\in V$. By the definition of tightness, $\calB'$ is tight \gls{wrt}\ $\calC$ by using $f$ itself. Therefore, $\calB' \leq \calB^*$ by the maximality of $B^*_1$. Consequently, $f\in W_{\calB^*,\calC}$.
\end{IEEEproof}

\section{Eccentricity of discrete sampling} \label[Appendix]{sec:ecc}

\begin{figure}[!htb]
  \centering
  \includegraphics[width=0.75\columnwidth]{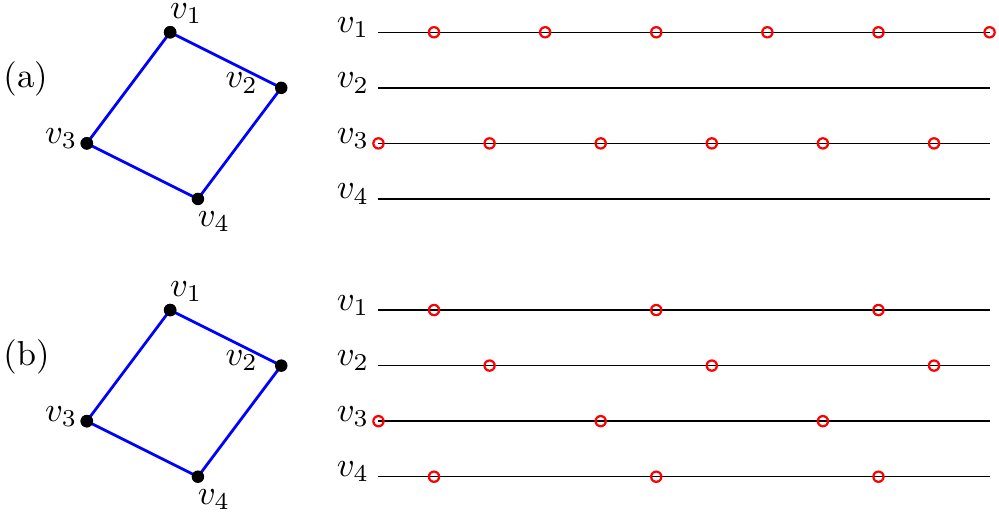}
  \caption{The graph consists of $4$ vertices $v_i,i=1,2,3,4$ forming an undirected cycle. Along each vertex $v_i$, the sample points are indicated on the right by the red circles. The sampling sets $S, S'$ of the schemes in (a) and (b) have the same sample rates. However, the eccentricity of $S'$ is only half of $S$. For $S$, we do not have any samples along $v_2$ and $v_4$, while for $S'$, we sample at a lower rate along $v_1$ and $v_3$.} \label{fig:cts1}
\end{figure}

In this appendix, we introduce the notion of eccentricity (illustrated in \cref{fig:cts1}) of discrete sampling to measure the ``shape'' of the sampling set.

\begin{Definition}
  Let $S$ be a discrete sampling set with sample rate $r(S)$ and $r_v$ for the sample rate for each $v\in V$, i.e., the sample rate of $S_v = S\cap \{v\}\times \mathbb{R}$ on $\mathbb{R}$. The \emph{eccentricity} $c(S)$ is defined as $$c(S) = |V|(\max_{v\in V} S_v)/r(S).$$
\end{Definition}

We follow the setup of \cref{sec:simple} on sampling for $\BWset$ with simple GFT bandwidth, which is more concrete. Let $V_0=\{v_1,\ldots, v_m\}, m=|V_0|$ be a minimal vertex set (c.f.\ Theorem~\ref{thm:feu}) for $\calB$ such that $\calB[v_i]\leq \calB[v_{i+1}], 1\leq i\leq m$. As in \cref{sec:simple}, for recovery of signals in $\BWset$, it suffices to sample at rate $2b_{v_i}$ along $v_i \in V_0$.

However, if $V_0$ is a small fraction of $V$ and $S$ only contains samples along $v\in V_0$, then the eccentricity $c(S)$ tends to be large. In this case, the bandwidths of a large proportion of $V$ are unused. This is not desirable in many applications. In fact, using the graph structure, it might be possible to reduce eccentricity without changing the sample rate.

\begin{Proposition} \label{prop:svo}
  Let $n=|V|$. Suppose $V^*$ of size $|V^*|=m'$ is a subset of vertices containing $V_0$ such that every $V'\subset V^*$ of size $m=|V_0|$ belongs to $\calU(\Lambda_0)$. Then there is a discrete sample set $S\subset V\times \mathbb{R}$ for perfect recovery of signals in $\BWset$ such that
  \begin{enumerate}[(1)]
    \item The sample rate $r(S)$ of $S$ is optimal, i.e., $r(S) = 2\sum_{v\in V_0}\calB[v]$.
    \item The eccentricity of $S$ is bounded by $$\frac{2n\calB[v_1]/[m'/m]+2n\sum_{i=1}^{m-1}(\calB[v_{i+1}]-\calB[v_i])/[(m'-i)/(m-i)]}{r(S)}.$$
  \end{enumerate}
\end{Proposition}

\begin{IEEEproof}
  First, we know that we may find a discrete sample set $S'$ for perfect recovery of $W_{\calB, \calC}$ by choosing sample rate $2\calB[v_i]$ along $v_i, 1\leq i\leq m$. By the non-uniform Nyquist-Shannon theorem, we may align $S'$ in such a way that for $1\leq i\leq m-1$, the $\mathbb{R}$ components of $S'_{v_i}$ is a subset of that of $S'_{v_{i+1}}$.

  We construct $S$ inductively. For the initial step, by the assumption on $V^*$, we may find $[m'/m]$ disjoint subset $V_1, \ldots, V_{[m'/m]}$ of $V^*$ such that sampling on each $V_j, 1\leq j\leq [m'/m]$ at any fixed $t\in \mathbb{R}$ has the same effect as sampling on $V_0$ at $t$. Therefore, the samples in $S'$ corresponding to $S'_{v_1}\times V_0$ can be replaced by samples along $V_1, \ldots, V_{[m'/m]}$ with rate $2\calB[v_1]/[m'/m]$ along each $V_j$. As a consequence, we can recover the restriction of any $f\in \BWset$ to $v_1$.

  Suppose we have dealt with $v_i$ so that we can recover the restriction of any $f\in \BWset$ to $V'=\{v_1,\ldots, v_i\}$, we proceed with $v_{i+1}$ similarly.  We only need to find $[m'-i/m-i]$ disjoint subsets $V_1, \ldots, V_{[(m'-i)/(m-i)]}$ of $V^*$ such that sampling on each $V_j \cup V'$ is equivalent to sampling on $V_0$. Therefore, we may re-distribute $(S'_{v_{i+1}}\backslash S'_{v_i})\times (V_0\backslash V')$ evenly along $V_j,1\leq j\leq [(m'-i)/(m-i)]$, by noting that at this stage the signal at each $v_i\in V'$ can already be perfectly recovered.

  By the construction, the eccentricity of $S$ is bounded by $$\frac{2n\calB[v_1]/[m'/m]+2n\sum_{i=1}^{m-1}(\calB[v_{i+1}]-\calB[v_i])/[(m'-i)/(m-i)]}{r(S)}.$$
\end{IEEEproof}

\bibliographystyle{IEEEtran}
\bibliography{IEEEabrv,StringDefinitions,allref}

\begin{thebibliography}{10}
\providecommand{\url}[1]{#1}
\csname url@samestyle\endcsname
\providecommand{\newblock}{\relax}
\providecommand{\bibinfo}[2]{#2}
\providecommand{\BIBentrySTDinterwordspacing}{\spaceskip=0pt\relax}
\providecommand{\BIBentryALTinterwordstretchfactor}{4}
\providecommand{\BIBentryALTinterwordspacing}{\spaceskip=\fontdimen2\font plus
\BIBentryALTinterwordstretchfactor\fontdimen3\font minus
  \fontdimen4\font\relax}
\providecommand{\BIBforeignlanguage}[2]{{%
\expandafter\ifx\csname l@#1\endcsname\relax
\typeout{** WARNING: IEEEtran.bst: No hyphenation pattern has been}%
\typeout{** loaded for the language `#1'. Using the pattern for}%
\typeout{** the default language instead.}%
\else
\language=\csname l@#1\endcsname
\fi
#2}}
\providecommand{\BIBdecl}{\relax}
\BIBdecl

\bibitem{Shu13}
D.~I. Shuman, S.~K. Narang, P.~Frossard, A.~Ortega, and P.~Vandergheynst, ``The
  emerging field of signal processing on graphs: Extending high-dimensional
  data analysis to networks and other irregular domains,'' \emph{IEEE Signal
  Process. Mag.}, vol.~30, no.~3, pp. 83--98, May 2013.

\bibitem{San13}
A.~Sandryhaila and J.~M.~F. Moura, ``Discrete signal processing on graphs,''
  \emph{IEEE Trans. Signal Process.}, vol.~61, no.~7, pp. 1644--1656, April
  2013.

\bibitem{San14}
------, ``Big data analysis with signal processing on graphs: Representation
  and processing of massive data sets with irregular structure,'' \emph{IEEE
  Signal Process. Mag.}, vol.~31, no.~5, pp. 80--90, Sept 2014.

\bibitem{Gad14}
A.~Gadde, A.~Anis, and A.~Ortega, ``Active semi-supervised learning using
  sampling theory for graph signals,'' in \emph{Proc. ACM SIGKDD Int. Conf. on
  Knowledge Discovery and Data Mining}, New York, NY, USA, 2014, pp. 492--501.

\bibitem{Don16}
X.~Dong, D.~Thanou, P.~Frossard, and P.~Vandergheynst, ``Learning {Laplacian}
  matrix in smooth graph signal representations,'' \emph{IEEE Trans. Signal
  Process.}, vol.~64, no.~23, pp. 6160--6173, Dec 2016.

\bibitem{Egi17}
H.~E. Egilmez, E.~Pavez, and A.~Ortega, ``Graph learning from data under
  {Laplacian} and structural constraints,'' \emph{IEEE J. Sel. Top. Signal
  Process.}, vol.~11, no.~6, pp. 825--841, Sept 2017.

\bibitem{Sha17}
R.~Shafipour, S.~Segarra, A.~G. Marques, and G.~Mateos, ``Network topology
  inference from non-stationary graph signals,'' in \emph{Proc. IEEE Int. Conf.
  Acoustics, Speech, and Signal Processing}, March 2017, pp. 5870--5874.

\bibitem{Gra18}
F.~Grassi, A.~Loukas, N.~Perraudin, and B.~Ricaud, ``A time-vertex signal
  processing framework: Scalable processing and meaningful representations for
  time-series on graphs,'' \emph{IEEE Trans. Signal Process.}, vol.~66, no.~3,
  pp. 817--829, Feb 2018.

\bibitem{Ort18}
A.~Ortega, P.~Frossard, J.~Kova\v{c}evi\'{c}, J.~M.~F. Moura, and
  P.~Vandergheynst, ``Graph signal processing: Overview, challenges, and
  applications,'' \emph{Proc. IEEE}, vol. 106, no.~5, pp. 808--828, May 2018.

\bibitem{Girault2018}
B.~{Girault}, A.~{Ortega}, and S.~S. {Narayanan}, ``Irregularity-aware graph
  fourier transforms,'' \emph{IEEE Transactions on Signal Processing}, vol.~66,
  no.~21, pp. 5746--5761, Nov 2018.

\bibitem{Ji19}
F.~Ji and W.~P. Tay, ``A {Hilbert} space theory of generalized graph signal
  processing,'' \emph{{IEEE} Trans. Signal Process.}, vol.~67, no.~24, pp. 6188
  -- 6203, Dec. 2019.

\bibitem{Aga13}
A.~{Agaskar} and Y.~M. {Lu}, ``A spectral graph uncertainty principle,''
  \emph{IEEE Trans. Inf. Theory}, vol.~59, no.~7, pp. 4338--4356, 2013.

\bibitem{Che15}
S.~{Chen}, R.~{Varma}, A.~{Sandryhaila}, and J.~{Kovačević}, ``Discrete
  signal processing on graphs: Sampling theory,'' \emph{{IEEE} Trans. Signal
  Process.}, vol.~63, no.~24, pp. 6510--6523, 2015.

\bibitem{Tsi15}
M.~{Tsitsvero}, S.~{Barbarossa}, and P.~{Di Lorenzo}, ``Signals on graphs:
  Uncertainty principle and sampling,'' \emph{{IEEE} Trans. Signal Process.},
  vol.~64, no.~18, pp. 4845--4860, 2016.

\bibitem{Mar16}
A.~G. {Marques}, S.~{Segarra}, G.~{Leus}, and A.~{Ribeiro}, ``Sampling of graph
  signals with successive local aggregations,'' \emph{{IEEE} Trans. Signal
  Process.}, vol.~64, no.~7, pp. 1832--1843, 2016.

\bibitem{Anis2016}
A.~{Anis}, A.~{Gadde}, and A.~{Ortega}, ``Efficient sampling set selection for
  bandlimited graph signals using graph spectral proxies,'' \emph{{IEEE} Trans.
  Signal Process.}, vol.~64, no.~14, pp. 3775--3789, 2016.

\bibitem{Yu19}
J.~{Yu}, X.~{Xie}, H.~{Feng}, and B.~{Hu}, ``On critical sampling of
  time-vertex graph signals,'' in \emph{2019 IEEE Global Conference on Signal
  and Information Processing (GlobalSIP)}, 2019, pp. 1--5.

\bibitem{Sha49}
C.~Shannon, ``Communication in the presence of noise,'' \emph{Proc. IRE},
  vol.~86, pp. 10--21, 1949.

\bibitem{Pesenson2008}
I.~Pesenson, ``Sampling in {Paley-Wiener} spaces on combinatorial graphs,''
  \emph{Trans. Amer. Math. Soc}, vol. 360, pp. 5603--5627, 2008.

\bibitem{Oxl92}
J.~Oxley, \emph{Matroid Theory}.\hskip 1em plus 0.5em minus 0.4em\relax Oxford
  University Press, 1992.

\bibitem{Ven04}
R.~Venkataramani and Y.~Bresler, ``Multiple-input multiple-output sampling:
  necessary density conditions,'' \emph{IEEE Trans. Inf. Theory}, vol.~50,
  no.~8, pp. 1754--1768, 2004.

\bibitem{Sid18}
A.~Sidford, \emph{Matroids and Maximum Flow}, 2018,
  \url{https://web.stanford.edu/class/cme305/Files/l4.pdf}.

\end{thebibliography}

\end{document}